\documentstyle[prl,aps,epsf]{revtex}
\begin{document}

\title{ Airy Distribution Function: From the Area Under a Brownian Excursion to the Maximal Height 
of Fluctuating Interfaces}
\author {Satya N. Majumdar $^{1,2}$ and Alain Comtet $^{2,3}$}
\address{
{\small $^1$Laboratoire de Physique Th\'eorique (UMR C5152 du CNRS), Universit\'e Paul
        Sabatier, 31062 Toulouse Cedex. France}\\
{\small $^2$Laboratoire de Physique Th\'eorique et Mod\`eles Statistiques,
        Universit\'e Paris-Sud. B\^at. 100. 91405 Orsay Cedex. France}\\
{\small $^3$Institut Henri Poincar\'e, 11 rue Pierre et Marie Curie, 75005 Paris, France}}

\maketitle

\date{\today}

\begin{abstract} 
The Airy distribution function describes the probability distribution of the area under a Brownian excursion
over a unit interval. Surprisingly, this function has appeared in a number of seemingly unrelated problems,
mostly in computer science and graph theory. In this paper, we show that this distribution function also appears
in a rather well studied physical system, namely the fluctuating interfaces. 
We present an exact solution for the distribution $P(h_m,L)$ of the maximal
height $h_m$ (measured with respect to the average spatial height) in
the steady state of a fluctuating interface
in a one dimensional
system of size $L$ with both periodic and free boundary conditions.
For the periodic case, we show that 
$P(h_m,L)=L^{-1/2}f\left(h_m L^{-1/2}\right)$
for all $L>0$ where the function $f(x)$ is the Airy distribution function. This result
is valid for both the Edwards-Wilkinson and the Kardar-Parisi-Zhang interfaces.  
For the free boundary case, the same scaling holds $P(h_m,L)=L^{-1/2}F\left(h_m L^{-1/2}\right)$, but
the scaling function $F(x)$ is different from that of the periodic case. We compute this scaling function explicitly
for the Edwards-Wilkinson interface and call it the $F$-Airy distribution function. 
Numerical simulations are in excellent agreement with our analytical results.
Our results provide a rather rare exactly solvable case for the distribution
of extremum of a set of {\it strongly correlated} random variables.
Some of these results were announced in a recent Letter [ S.N. Majumdar and A. Comtet, Phys. Rev. Lett., {\bf 92}, 225501 
(2004) ].

\vskip 5mm 
\noindent PACS numbers: 02.50.-r, 89.75.Hc, 89.20.Ff 
\end{abstract}


\section{Introduction}

A Brownian excursion $x(\tau)$ is a conditioned one dimensional Brownian motion over the time 
interval $0\le \tau\le T$ such that its path starts and ends at the origin $x(0)=x(T)=0$, but is constrained to 
stay
positive in between (see Fig. 2). The area under the excursion, $A=\int_0^T x(\tau) d\tau$, is clearly a random 
variable taking a different value for each realization of the excursion. A natural question that the 
mathematicians
have studied quite extensively over the past two decades is: what is the probability distribution  
$P(A,T)$ of the area
under a Brownian excursion over the interval $[0,T]$? Since the typical lateral displacement of the excursion
at time $\tau$ scales as $\sqrt{\tau}$, it follows that one can trivially rescale $x(\tau)=\sqrt{\tau} y(\tau/T)$
where $y(u)$ represents a Brownian excursion over the unit interval $u\in [0,1]$. Hence,
the area $A=\int_0^T x(\tau) d\tau\equiv T^{3/2}\int_0^{1} y(u)du $ 
scales as $T^{3/2}$. Thus the area distribution has a scaling form, $P(A,T)= T^{-3/2} 
f\left(A/T^{3/2} 
\right)$. The normalization condition $\int_0^{\infty} P(A,T)dA=1$ demands a prefactor $T^{-3/2}$
and also the conditions: $f(x)\ge 0$ for all $x$ and  $\int_0^{\infty} f(x)dx=1$. One then interprets the 
scaling function $f(x)$ as the 
distribution of the 
area under the Brownian excursion $y(u)$ over a {\em unit} interval $u\in [0,1]$. 
The function $f(x)$, or rather its Laplace transform, was first computed analytically
by Darling\cite{Darling} and Louchard\cite{Louchard},
\begin{equation}
{\tilde f} (s)= \int_0^{\infty} f(x) e^{-sx} dx= s\sqrt{2\pi}\sum_{k=1}^{\infty} e^{-\alpha_k s^{2/3} 
2^{-1/3}},
\label{airy1}
\end{equation}
where $\alpha_k$'s are the magnitudes of the zeros of the standard Airy function $Ai(z)$ on the negative real 
axis. For example, $\alpha_1=2.3381\dots$, $\alpha_2=4.0879\dots$, $\alpha_3=5.5205\dots$ etc. Since the 
expression of $f(x)$ involves the zeros of Airy function, the function $f(x)$ has been named the Airy 
distribution function, which should not be confused with the Airy function $Ai(x)$ itself.
Even though the Eq. (\ref{airy1}) provides a formally exact expression of the Laplace transform, it turns out 
the calculation of the moments $M_n= \int_0^{\infty} x^n f(x)dx$ is highly nontrivial and they can be 
determined only recursively\cite{Takacs} (see Section II). Takacs was also able to formally invert the Laplace transform 
in Eq. (\ref{airy1}) to obtain\cite{Takacs},
\begin{equation}
f(x)= {{2\sqrt{6}}\over {x^{10/3}}}\sum_{k=1}^{\infty} e^{-b_k/x^2} b_k^{2/3}
U(-5/6, 4/3, b_k/x^2),
\label{fx1}
\end{equation}
where $b_k= 2\alpha_k^3/{27}$ and $U(a,b,z)$ is the confluent hypergeometric function\cite{AS}.
The function $f(x)$ has the asymptotic tails\cite{Takacs,CSY},
\begin{eqnarray}
f(x) &\sim & x^{-5}\, e^{- 2\alpha_1^3/{27 x^2}} \quad {\rm as}\quad x\to 0 \nonumber \\
f(x) &\sim & e^{-6 x^2} \quad {\rm as}\quad x\to \infty.
\label{asymfx}
\end{eqnarray}
A plot of this function, obtained by evaluating the sum in Eq. (\ref{fx1}) using the Mathematica, is provided in Fig. 1.
\begin{figure}[htbp]
\epsfxsize=8cm
\centerline{\epsfbox{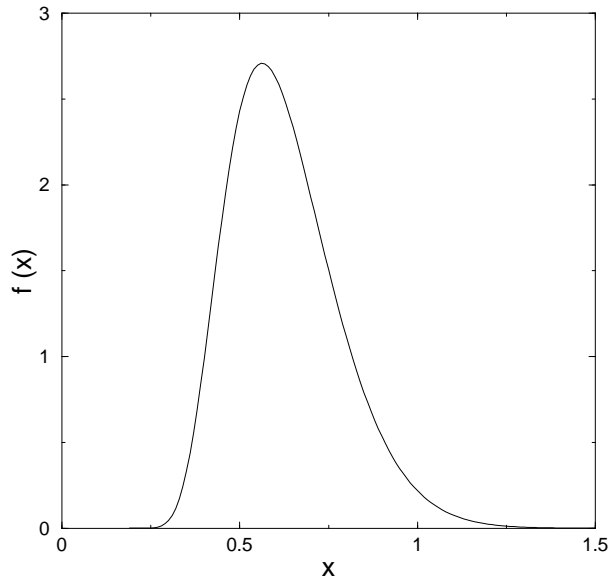}}
\caption{A Mathematica plot of the Airy distribution function $f(x)$ in Eq. (\ref{fx1}).}
\label{fig:airypbc}
\end{figure}

So, why would anyone care about such a complicated function? The reason behind the sustained interest
and study\cite{Takacs,CSY,FPV,FL,FSS} of this function $f(x)$ seems to be the fact that it keeps 
resurfacing in a number of seemingly unrelated problems, mostly in computer science and graph theory [for a 
list of such 
problems see \cite{FPV}]. For example, the function $f(x)$ describes the distribution of the cost of the construction
of a linear table for data storage using the linear probing with random
hashing algorithm\cite{FPV}. The 
function $f(x)$ also describes the distribution of the total path
length in Catalan trees\cite{Takacs}. The generating 
function for the number of inversions in trees involves the Airy distribution function $f(x)$\cite{MR}.
Also, the moments $M_n$'s of the function $f(x)$ appear in the enumeration of the connected components in
a random graph\cite{Wright,FKP}. Recently, it has been conjectured and subsequently tested numerically
that the asymptotic distribution of the area of two dimensional self-avoiding polygons is also given by the 
Airy distribution function $f(x)$\cite{RGJ}. Besides, numerical evidence suggests that the area enclosed 
by the outer boundary of planar random 
loops is also distributed according to the Airy distribution function $f(x)$\cite{R1}. 
Given the widespread occurrence of this function $f(x)$ in various combinatorial problems, it is natural to ask 
whether this function also appears
in a more realistic physical system. In a recent Letter\cite{MC1}, we have demonstrated that indeed this Airy 
distribution function also appears in a rather well studied physical system, namely the one dimensional 
fluctuating interfaces.

Fluctuating interfaces have been widely studied over the last two decades as they appear in a variety of 
physical systems such as growing crystals, molecular beam epitaxy, fluctuating steps on metals and growing 
bacterial colonies\cite{Review,HZ}. The past theoretical efforts mostly concentrated in characterizing the 
universal scaling properties of the surface roughness measured via the average width of the surface.
This average roughness, however, fails to measure the possible extreme fluctuations which, though rare, may 
play an important role under many situations. For example, extreme fluctuations play a significant role in 
batteries where a short circuit may occur if the highest point of the metal surface on one electrode reaches 
the one on the opposite electrode. There is also a compelling theoretical motivation for studying 
the extreme fluctuations. While the extreme value statistics is well
understood for a set of {\em independent} random variables\cite{Extreme}, it becomes nontrivial
when the random variables are {\em correlated}. The extreme value problems associated with correlated variables
have appeared recently in a variety of problems
ranging from
disordered systems\cite{DS} to a number of
computer science problems that include the growing search trees\cite{Trees} and the networks\cite{Net}.
A fluctuating interface is an ideal candidate for studying the extreme statistics of correlated 
variables since the 
heights at two different points on the interface are correlated. A
knowledge of the distribution of their maximum (or minimum) may possibly provide
important insights into this important general class of extreme value problems
of correlated variables.
 
Motivated by these observations, Raychaudhury et. 
al.\cite{RCPS} recently proposed to study the statistics of 
the global maximal relative height (MRH) (measured with respect to the spatially averaged growing height)
of a fluctuating $(1+1)$-dimensional interface. By $(1+1)$-dimensional interface one means
an interface characterized by a scalar height function $H(x,t)$ growing over a linear substrate.
For a substrate of finite size $L$, the
system reaches a stationary state in the long time limit. In this stationary state, the MRH $h_m$ is expected to 
scale as the surface roughness, $h_m\sim L^{\alpha}$ for large $L$, where $\alpha$ is the roughness exponent.
This suggests, quite generically, a scaling form for the normalized probability distribution of the MRH, $P(h_m,L)\sim 
L^{-\alpha} f_1\left(h_m/L^{\alpha}\right)$ where $f_1(x)$ is a scaling function.
This was demonstrated 
numerically in Ref. \cite{RCPS}   
for a one dimensional lattice model belonging to the Edwards-Wilkinson (EW) universality class\cite{Edwards}.     
Further, it was argued that the scaling function $f_1(x)$ is sensitive to the boundary conditions\cite{RCPS}.

In our previous Letter \cite{MC1}, using path integral techniques, we presented an exact solution of the 
stationary MRH distribution $P(h_m, L)$ for the $(1+1)$-dimensional EW model, with both periodic and free
boundary conditions. For the periodic case, we showed that $P(h_m,L)=L^{-1/2}\,f\left(h_m L^{-1/2}\right)$
for all $L>0$ where the scaling function $f(x)$ is precisely the Airy distribution function defined in Eq. 
(\ref{airy1}). For the free boundary case, the same scaling was found to hold though
the scaling function was different from that of the periodic case. 
The purpose of the present paper is twofold: (i) to provide detailed derivations of various results that were 
just announced in Ref.\cite{MC1} and (ii) to extend our results to other systems such as the
Kardar-Parisi-Zhang (KPZ) interface\cite{KPZ}. Our main results, along with a layout of the paper, are
summarized below.
\begin{enumerate}
\item In Section II-A, we provide a new physical derivation, using methods of path 
integral, of the 
distribution of the area under a Brownian excursion. This result is used later in Sec. IV-A 
in the context of the MRH distribution of an EW interface with periodic boundary condition. In 
Section II-B, we derive the distribution of the area under a Brownian meander (a Brownian
meander is a one dimension Brownian motion $x(\tau)$ over the time interval $0\le \tau \le T$ that 
is pinned to the origin at one end $x(0)=0$, but is free at the other end at $\tau=T$, 
and is constrained to stay positive in between). 
These results for the Brownian meander will be used later to discuss the MRH distribution
of an EW interface with free boundary condition in Sec. IV-B. 
The Section II can be read independently of the rest of the paper and the results here are interesting
by themselves.

\item Section III includes a detailed derivation of various correlation functions in the EW model in
two opposite regimes: (a) {\bf growing regime } where the correlation length
of the interface $\xi(t)\sim t^{1/z}$ ($z$ being the dynamical exponent) is less than the system size,
$\xi(t)<< L$ and (b) {\bf stationary regime} when $\xi(t)>> L$ and thus the joint probablity distribution of 
height fluctuations become time independent. An explicit form of this joint distribution in the stationary regime
is derived for both periodic and free boundary conditions. 

\item In Section IV, we derive analytically the stationary MRH distribution in $(1+1)$-dimensional
EW model. 
Section IV-A discusses the periodic boundary condition case, where the scaled MRH distribution
is shown to be exactly the Airy distribution function defined in Eq. (\ref{airy1}). The moments of this
distribution are also computed exactly. The analytical results are then compared to
the numerical results obtained from the direct numerical integration of the EW equation and an excellent agreement
is found. In Section IV-B, we derive the MRH distribution for the EW interface with free boundary 
condition and show that it is related to the  area under two independent
Brownian meanders. We call the associated scaling function $F$-Airy distribution function where $F$ refers to the
free boundary condition. 
We also compute the moments and the asymptotic properties of this distribution analytically. As in the periodic case, we 
find excellent agreement between the analytical and the numerical results.

\item In Section V, we discuss the $(1+1)$-dimensional KPZ equation and show that for the periodic boundary 
condition, the stationary MRH distribution is again given by the Airy 
distribution function in Eq. (\ref{airy1}). This is further confirmed by direct numerical integration of
the $(1+1)$-dimensional KPZ equation. 

\item Section VI discusses the MRH distribution in the growing regime $t<< L^z$. Using the well
known theory of the extreme value statistics of {\em independent} random variables, we show that
the appropriately scaled MRH distribution is universal and is given by the Gumbel function.

\item We conclude in Section VII with a summary, experimental consequences of our results and some open questions. 

\end{enumerate}

\section{The Area under a Brownian Excursion and a Brownian Meander}

In this section, we provide a new derivation, using path integral methods, of the distribution of 
the area under a Brownian excursion and a Brownian meander. These results were originally derived by 
mathematicians\cite{Darling,Louchard,Takacs,PW,Jeanblanc} using various probabilistic methods.
The path integral method presented here, though perhaps not rigorous in the strict mathematical sense, does 
provide a simpler physical derivation. Thus the results of this section are interesting by themselves
and can be read independently of the rest of the paper that discusses the fluctuating interfaces. This 
section will also serve as an introduction to the path integral method to be used in later sections to study the
maximal height fluctuations in interfaces.

\subsection{Brownian Excursion}

A Brownian excursion is a one dimensional Brownian motion
over the time interval $0\le \tau \le T$ that starts and ends at the origin $x(0)=x(T)=0$, but is 
constrained to stay positive in between. We are interested in computing the probability distribution 
$P(A,T)$ of the area $A=\int_0^{T} x(\tau)d\tau$ under the excursion. The Brownian 
excursion, as described above, is however a bit ill defined for a Brownian walk in 
continuous space and time. For such a walk, it is well known that if the walker crosses zero once, it
recrosses the zero infinitely many times immediately after the first crossing. 
Thus, for a continuous space-time Brownian motion, it is impossible to enforce the constraint $x(0)=0$ and 
simultaneously 
forcing it to stay positive immediately after. There are two ways to go around this problem. The cleanest way to study 
an excursion is to consider 
a discrete time random walk moving on a discrete one dimensional lattice and then sample all configurations that 
start at the origin
and come back to the origin for the first time after an even $T=2n$ number of steps. Such paths are called 
Dyck paths\cite{Takacs}. One then defines the area under such a path as 
$A=\sum_{i=1}^{2n} x_i$ where $x_i$ is the position of the walker at step $i$.
By appropriately taking the continuum 
limit, one would then arrive at the Brownian excursion\cite{NThe}. This method, though conceptually 
clear, is somewhat cumbersome mathematically. 

Here we devise a second method that turns out to be more amenable to path 
integral techniques. We consider the Brownian motion in continuous space and time, but introduce 
a small cut-off $\epsilon$ by hand as shown in Fig. 2. More precisely, our Brownian motion
starts at $x(0)=\epsilon$ and comes back after time $T$ to $x(T)=\epsilon$, without crossing
the origin in between. We will first derive the probability density $P(A,T, \epsilon)$ of the area
$A=\int_0^{T} x(\tau)d\tau$ for a fixed $\epsilon$ and then finally take the limit $\epsilon\to 0$.
We will show that the limiting distribution exists and is precisely the Airy distribution function 
defined in Eq. (\ref{airy1}).
\begin{figure}[htbp]
\epsfxsize=8cm
\centerline{\epsfbox{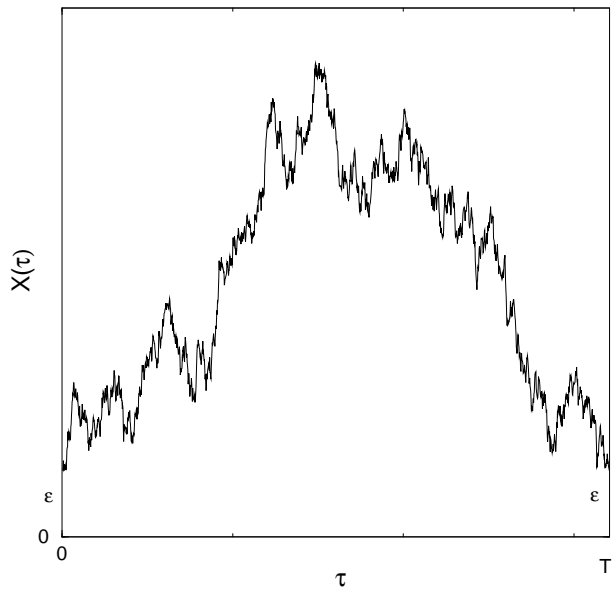}}
\caption{A Brownian excursion over the time interval $0\le \tau \le T$ starting at
$x(0)=\epsilon$ and ending at $x(T)=\epsilon$ and staying positive in between.}
\label{fig:excursion}
\end{figure}

The probability measure associated with an unconstrained Brownian path $x(\tau)$ over the time 
interval $0\le \tau \le T$ is proportional to
$\exp\left[-\frac{1}{2} \int_0^{T} \left(\frac{dx}{d\tau}\right)^2 d\tau\right]$. In addition, we need 
to incorporate the constraint that it stays positive between $0$ and $T$ which can be achieved by 
multiplying the above measure with the indicator function $\prod_{\tau=0}^{T} 
\theta\left[x(\tau)\right]$ which is $1$ if the path stays positive in $0\le \tau\le T$ and zero
otherwise. The distribution $P(A,T)$ of the area
$A=\int_0^T x(\tau)d\tau$ under this Brownian excursion can then be expressed as a path integral
\begin{equation}
P(A,T)= \frac{1}{Z_E}\int_{x(0)=\epsilon}^{x(T)=\epsilon} {\cal D}x(\tau)\,e^{-{1\over 
{2}}\int_0^{T}d\tau\, (dx/{d\tau})^2}\,\prod_{\tau=0}^{T}
\theta\left[x(\tau)\right]\, \delta\left(\int_0^T x(\tau)d\tau -A\right).
\label{exarea1}
\end{equation}
Note that we have suppressed the $\epsilon$ dependence of $P(A,T)$ for brevity.
The normalization of the distribution, $\int_0^{\infty} P(A,T)dA=1$, is ensured by the following 
definition
of the partition function $Z_E$ of the Brownian excursion
\begin{equation}
Z_E= \int_{x(0)=\epsilon}^{x(T)=\epsilon} {\cal D}x(\tau)\,e^{-{1\over
{2}}\int_0^{T}d\tau\, (dx/{d\tau})^2}\, \prod_{\tau=0}^{T}
\theta\left[x(\tau)\right].
\label{expart1}
\end{equation}
All paths inside the path integrals in Eqs. (\ref{exarea1}) and (\ref{expart1}) propagate
from their initial value $x(0)=\epsilon$ at $\tau=0$ to their final value $x(0)=\epsilon$ at $\tau=T$.

We first evaluate the partition function $Z_E$. This is easy since one can identify
the quantity inside the exponential in Eq. (\ref{expart1}) as the action corresponding
to a single particle quantum Hamiltonian, ${\hat H}_0\equiv -\frac{1}{2} \frac{d^2}{dx^2} +V_0(x)$,
where the potential $V_0(x)=0$ for $x>0$ and $V_0(x)=\infty$ for $x\le 0$. The infinite potential
for $x\le 0$ ensures that the path never crosses zero and thus takes care of the indicator
function $\prod_{\tau=0}^{T}\theta\left[x(\tau)\right]$ in Eq. (\ref{expart1}). Using the standard 
bra-ket 
notation, the partition function $Z_E$ is then simply the propagator
\begin{equation}
Z_E= <\epsilon|e^{-{\hat H}_0 T}| \epsilon>.
\label{exprop1}
\end{equation}
It is easy to see that the Hamiltonian ${\hat H}_0$ has continuous eigenvalues $E_0=k^2/2$ labelled by $k\ge 0$
and the corresponding eigenfunctions that vanish at the origin are given by, $\psi_k(x)= 
\sqrt{\frac{2}{\pi}}\sin (kx)$. Expanding the propagator in the energy eigenbasis, one gets
\begin{equation}
Z_E = \int_0^{\infty} dk |\psi_k(\epsilon)|^2 e^{-k^2T/2}= \frac{1}{\sqrt{2\pi 
T}}\left(1-e^{-2\epsilon^2/T}\right).
\label{expart2}
\end{equation}
Note that $Z_E$ is just the probability that a Brownain motion goes from $x(0)=\epsilon$ to 
$x(T)=\epsilon$ in time $T$ without crossing the origin, and hence can also be computed 
by the standard image method which yields the same answer\cite{Redner} as in Eq. (\ref{expart2}).

We now turn to the evalution of $P(A,T)$ in Eq. (\ref{exarea1}). Taking the Laplace
transform, ${\tilde P}(\lambda, T)= \int_0^{\infty} P(A,T) e^{-\lambda A} dA$
where $\lambda\ge 0$, in Eq. (\ref{exarea1}) gives,
\begin{equation}
{\tilde P}(\lambda,T)= \frac{1}{Z_E}\int_{x(0)=\epsilon}^{x(T)=\epsilon} {\cal D}x(\tau)\,
e^{- \int_0^{T}d\tau\, \left[ \frac{1}{2}\left(dx/{d\tau}\right)^2 +\lambda 
x(\tau)\right]} 
\prod_{\tau=0}^{T}\theta\left[x(\tau)\right].
\label{exarea2}
\end{equation}
The numerator in Eq. (\ref{exarea2}) can then be identified as the propagator 
$<\epsilon|e^{-{\hat H}_1 T}|\epsilon>$,  with the Hamiltonian 
${\hat H}_1\equiv -\frac{1}{2} \frac{d^2}{dx^2} +V_1(x)$
where $V_1(x)$ is a triangular potential: $V_1(x)= \lambda x$ for $x>0$ and $V_1(x)=\infty$ for $x\le 
0$. The later condition again takes care of the fact that the paths do not cross zero. The Hamiltonian
${\hat H}_1$ has only bound states and hence discrete eigenvalues. Solving the corresponding 
Schr\"odinger equation, one finds that the wavefunction (up to a normalization constant), that vanishes 
as $x\to \infty$, is simply given 
by, $\psi_E(x)= Ai \left[ (2\lambda)^{1/3}\left(x- E/\lambda\right)\right]$, where $Ai(z)$ is the 
standard Airy function\cite{AS}. The condition that the 
wavefunction should vanish at $x=0$ determines the discrete eigenvalues, $E_k= \alpha_k \lambda^{2/3} 
2^{-1/3}$ for $k=1,2,\ldots$, where $\alpha_k$'s are the magnitude of the zeros of $Ai(z)$   
on the negative real axis. For example, one has $\alpha_1=2.3381\dots$, $\alpha_2=4.0879\dots$, 
$\alpha_3=5.5205 \dots $ etc. Thus the normalized eigenfunction is given by
\begin{equation}
\psi_k(x)= \frac{Ai\left[(2\lambda)^{1/3}x - \alpha_k\right]}{\sqrt{\int_0^{\infty} 
Ai^2\left[(2\lambda)^{1/3}y-\alpha_k\right]dy}}.
\label{exeigen1}
\end{equation}
The integral $\int_0^{\infty}
Ai^2\left[(2\lambda)^{1/3}y-\alpha_k\right]dy= (2\lambda)^{-1/3}\int_{-\alpha_k}^{\infty} Ai^2(z)dz$
can be further simplified by using the identity, 
$\int_{-\alpha_k}^{\infty}Ai^2(z)dz= [Ai'(-\alpha_k)]^2$\cite{Albright} where $Ai'(z)= dAi(z)/dz$.
Expanding the propagator $<\epsilon|e^{-\hat H_1 T}|\epsilon>$ into the energy eigenbasis we get
\begin{equation}
<\epsilon|e^{-\hat H_1 T}|\epsilon>= \sum_{k=1}^{\infty} |\psi_k(\epsilon)|^2 e^{-\lambda^{2/3}2^{-1/3} 
\alpha_k T}.
\label{exprop2}
\end{equation}
Using the exact eigenfunctions from Eq. (\ref{exeigen1}) and then substituting all the expressions
in Eq. (\ref{exarea2}) we get
\begin{equation}
{\tilde P}(\lambda,T)=\frac{(2\lambda)^{1/3}\,\sqrt{2\pi T}}{\left(1-e^{-2\epsilon^2/T}\right)}\,
\sum_{k=1}^{\infty} 
\frac{Ai^2\left[(2\lambda)^{1/3}\epsilon-\alpha_k\right]}{{Ai'}^2[-\alpha_k]}\, e^{-\lambda^{2/3}2^{-1/3} 
\alpha_k T}.
\label{exarea3}
\end{equation} 
We are now ready to take the $\epsilon\to 0$ limit. Expanding both the numerator and the denominator
in Eq. (\ref{exarea3}) in a Taylor series in $\epsilon$ and taking the $\epsilon\to 0$ limit, we
get a simplified result, 
\begin{equation}
{\tilde P}(\lambda,T)= \sqrt{2\pi}\, (\lambda T^{3/2})\sum_{k=1}^{\infty} e^{-2^{-1/3}\alpha_k
(\lambda T^{3/2})^{2/3}}.
\label{exarea4}
\end{equation}
Note that the right hand side of Eq. (\ref{exarea4}) is a function of only one scaling
combination $s=\lambda T^{3/2}$. This demands that the distribution $P(A,T)$ must have the scaling form
$P(A,T)= T^{-3/2}\,f\left(A\,T^{-3/2}\right)$, so that its Laplace transform is a function
of only the combination $s=\lambda T^{3/2}$,
${\tilde P}(\lambda, T)= \int_0^{\infty} f(x) e^{-\lambda T^{3/2} x} dx$. Comparing this
to the right hand side of Eq. (\ref{exarea4}) we arrive at the final result
\begin{equation}
{\tilde f}(s)= \int_0^{\infty} f(x)e^{-sx}dx = s\sqrt{2\pi} \sum_{k=1}^{\infty} e^{-\alpha_k 
s^{2/3}2^{-1/3}}.
\label{exarea5}
\end{equation}

{\bf Moments of the area distribution:} Using the scaling form, $P(A,T)= 
T^{-3/2}\,f\left(A\,T^{-3/2}\right)$, one finds that the moments $A_n= \int_0^{\infty} A^n 
P(A,T) dA= M_n T^{3n/2}$, where $M_n= \int_0^{\infty} f(x)\, x^n dx$ are the moments of the area
under the excursion over a unit interval. The extraction of the moments $M_n$ explicitly from the 
Laplace transform in Eq. 
(\ref{exarea5}) is highly nontrivial. However, starting from Eq. (\ref{exarea5}), Takacs found
a recursive method to compute the moments\cite{Takacs}. We summarize the main results here without the details.
One first defines a set of new variables $K_n$ via the relation
\begin{equation}
M_n= \sqrt{\pi}\, 2^{(4-n)/2}\, \frac{\Gamma(n+1)}{\Gamma\left(\frac{3n-1}{2}\right)}\,K_n,
\label{kn1}
\end{equation}
where $\Gamma(x)$ is the Gamma function. The variables $K_n$'s subsequently satisfy
a nonlinear recurrence 
\begin{equation}
K_n = \frac{3n-4}{4} K_{n-1} + \sum_{j=1}^{n-1} K_j K_{n-j},
\label{exmoment1}
\end{equation}
starting with $K_0=-1/2$. This produces recursively the exact values of the moments $M_n$ for any $n$. For example,
the first few values are, 
\begin{equation}
M_0=1, \quad\quad M_1=\frac{1}{2}\sqrt{\frac{\pi}{2}}, \quad\quad
M_2=\frac{5}{12}, \quad\quad  M_3= \frac{15}{64}\sqrt{\frac{\pi}{2}}, \quad\quad M_4= \frac{221}{1008}, \quad \ldots
\label{exmoment2}
\end{equation}
These moments will be used later in Sec. IV-A in the context of the maximal height fluctuations in interfaces with 
periodic boundary conditions.

\subsection{Brownian Meander}   
 
A Brownian meander is a path of a one dimensional Brownian motion over the time interval $0\le \tau\le T$
that starts at the origin $x(0)=0$ and stays positive upto $\tau=T$. The difference between an excursion 
and a meander is that in the former case, the path comes back to the origin at the end of the interval,
$x(T)=0$. In the case of the meander, the Brownian walker is free to arrive at any final position as long as 
the final 
position is positive. As in the case of the Brownian excursion in the previous subsection, we consider
the Brownian meander in continuous space and time, but introduce a small cut-off in the initial position
$x(0)=\epsilon$, and eventually take the $\epsilon\to 0$ limit. In Fig. 3, we show
a Brownian meander that finally arrives at the position $b>0$ at time $\tau=T$, starting at
$x(0)=\epsilon$. 
Contrary to the Brownian excursion where the final position is same
as the initial position i.e., $x(T)=b=\epsilon$, for the meander we need to integrate the paths over all
possible final positions $b>0$. 
\begin{figure}[htbp]
\epsfxsize=8cm
\centerline{\epsfbox{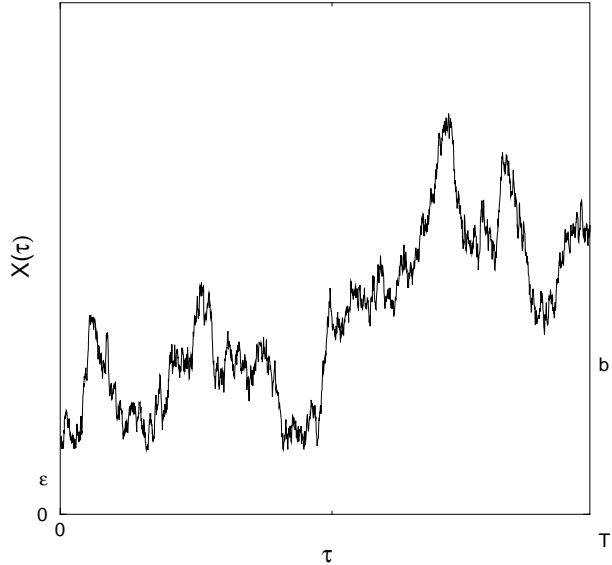}}
\caption{A Brownian meander over the time interval $0\le \tau \le T$ starting at
$x(0)=\epsilon$ and ending at $x(T)=b$ and staying positive in between.}
\label{fig:meander}
\end{figure}

As before, we are interested in computing the probability 
distribution $P(A,T)$ of the area $A= \int_0^T x(\tau)d\tau$ under a meander. 
We suppress the $\epsilon$ and $b$ dependence of $P(A,T)$ for brevity.
Following the same 
arguments as in Sec. II-A, one can easily write down an expression for the 
Laplace transform, ${\tilde P}(\lambda, T)= \int_0^{\infty} P(A,T) e^{-\lambda A} dA$, as a ratio of 
two 
propagators,
\begin{equation}
{\tilde P}(\lambda, T)= \frac{\int_0^{\infty}db <b|e^{-{\hat H}_1 T}|\epsilon>}{\int_0^{\infty}db 
<b|e^{-{\hat H}_0 T}|\epsilon>} \, ,
\label{marea1}
\end{equation}
where the Hamiltonians ${\hat H}_0$ and ${\hat H}_1$ are the same as in the previous subsection.

The denominator, $Z_M= \int_0^{\infty} db
<b| e^{-{\hat H}_0 T}|\epsilon>$, is the partition function of the Brownian meander and can be easily 
evaluated by decomposing the propagator in the energy eigenbasis of ${\hat H}_0$. We get
\begin{equation}
Z_M = {\rm erf}(\epsilon/{\sqrt{2T}}),
\label{mpart1}
\end{equation}
where ${\rm erf}(x)= \frac{2}{\sqrt{\pi}}\int_0^{x} e^{-u^2}du$ is the standard Error function.
Note that $Z_M$ is simply the probability that a Brownian walker does not return to the origin upto time 
$T$ starting at the initial position $\epsilon$ and hence, can be computed also via the method of 
images\cite{Redner}.

The propagator in the numerator in Eq. (\ref{marea1}) can be evaluated by 
expanding it in the energy eigenbasis of ${\hat H}_1$, which were detailed in sec. II-A.
We do not repeat the calculations here and just mention the final result. We get
\begin{equation}
{\tilde P}(\lambda, T)= \frac{1}{{\rm erf}({\epsilon}/{\sqrt{2T}})}\,
\sum_{k=1}^{\infty} \frac{ Ai[(2\lambda)^{1/3} \epsilon -\alpha_k]\, \int_{-\alpha_k}^{\infty} Ai(z)dz}
{{Ai'}^2(-\alpha_k)} e^{-\lambda^{2/3}2^{-1/3}
\alpha_k T},
\label{marea2}
\end{equation}
where $\alpha_k$'s are the magnitude of the zeros of the Airy function. 
Taking the $\epsilon\to 0$ limit we obtain a simpler expression
\begin{equation}
{\tilde P}(\lambda,T)= \sqrt{\pi}\, 2^{-1/6}\, (\lambda T^{3/2})^{1/3}\,\sum_{k=1}^{\infty} B(\alpha_k) 
e^{-2^{-1/3}\alpha_k
(\lambda T^{3/2})^{2/3}},
\label{marea3}
\end{equation}
where 
\begin{equation}
B(\alpha_k)= \frac{\int_{-\alpha_k}^{\infty}Ai(z)dz}{Ai'(-\alpha_k)}.
\label{balphak}
\end{equation} 
One should compare this 
result with its analogue in the case of the Brownian excursion in Eq. (\ref{exarea4}). As in the case 
of the excursion, the right hand side of Eq. (\ref{marea3}) is a function of only one scaling combination,
$s=\lambda T^{3/2}$. This indicates that $P(A, T)= T^{-3/2}\,g\left(A\,T^{-3/2}\right)$, where the 
Laplace transform of the scaling function $g(x)$ is obtained from Eq. (\ref{marea3}),
\begin{equation}
{\tilde g}(s)= \int_0^{\infty} g(x)\, e^{-sx}dx = \sqrt{\pi}\, 2^{-1/6}\, s^{1/3}\,\sum_{k=1}^{\infty} 
B(\alpha_k) e^{-\alpha_k s^{2/3}2^{-1/3}}. 
\label{marea4}
\end{equation}

To make contact with the existing results in the probability literature, we first note that
the Laplace transform, ${\tilde P}(\lambda,T)=\int_0^{\infty} P(A,T) e^{-\lambda A}dA$ can be 
regarded as the expectation value 
$E[e^{-\lambda A}]$ with respect to the probability distribution $P(A,T)$. Next, in terms
of a new scaling variable $t= 2^{-1/3}\lambda^{2/3} T$, Eq. (\ref{marea3}) can be rewritten in a simpler form,
\begin{equation}
E\left[e^{- \sqrt{2} t^{3/2} a} \right]= \sqrt{\pi t}\, \sum_{k=1}^{\infty} B(\alpha_k) 
e^{-\alpha_k t},
\label{marea5}
\end{equation}
where we have defined the scaled area $a= A/T^{3/2}$. Note that, as in the case of an excursion, the interval length $T$
only appears as a trivial scale factor and hence 
one can interpret $a$
as the area under a Brownian meander over the unit interval $[0,1]$.
Dividing both sides of Eq. (\ref{marea5}) by $\sqrt{\pi t}$ and taking a further Laplace transform 
with respect to $t$, we get
\begin{equation}
\int_0^{\infty} \frac{dt}{\sqrt{\pi t}}e^{-ut} E\left[e^{- \sqrt{2} t^{3/2} a}\right]= 
\sum_{k=1}^{\infty} \frac{B(\alpha_k)}{(\alpha_k+u)}= \frac{ \int_u^{\infty} Ai(z)dz}{Ai(u)}.
\label{marea6}
\end{equation}
The second equality in Eq. (\ref{marea6}) can be proved by replacing the sum by a contour integration
in the complex plane, a derivation of which is given in Appendix-A.
This result in Eq. (\ref{marea6}) is identical to an expression derived by Perman and 
Wellner\cite{PW} using probabilistic methods. 
Here we have provided a simpler physical derivation using path integral
techniques.

{\bf Moments of the area distribution:} The scaling form 
$P(A,T)= T^{-3/2}\,g\left(A\,T^{-3/2}\right)$ indicates that the moments of the area  
under a Brownian meander are given as
$A_n = \int_0^{\infty} A^n P(A, T) dA= a_n T^{3n/2}$, where
$a_n= \int_0^{\infty}g(x)\, x^n  dx$ are the moments of the area under a Brownian meander over the unit interval. 
As in the case of Brownian excursion, the extraction of
the moments $a_n$'s explicitly from the Laplace transform
in Eq. (\ref{marea4}) is not easy. However, starting
from the alternative expression of the Laplace transform
in Eq. (\ref{marea6}), Perman and Wellner were able to obtain the moments $a_n$'s recursively\cite{PW}. Here we 
briefly summarize their results without the details. We first define a set of new variables $R_n$'s via the relation
\begin{equation}
a_n= \sqrt{\pi}\, 2^{-n/2}\, \frac{\Gamma(n+1)}{\Gamma\left(\frac{3n+1}{2}\right)}\,R_n.
\label{rn1}
\end{equation}
Next, the $R_n$'s were found to satisfy the following recursion relations\cite{PW} for all $n\ge 1$,
\begin{eqnarray}
R_n &=& \beta_n- \sum_{j=1}^n \gamma_j R_{n-j} \nonumber \\
\beta_n &=& \gamma_n + \frac{3}{4}(2n-1) \beta_{n-1} \nonumber \\
\gamma_n &=& \frac{(36)^{-n}}{\Gamma(n+1)}\,\frac{\Gamma\left(3n+1/2\right)}{\Gamma\left(n+1/2\right)}.
\label{rn2}
\end{eqnarray}
Using Eqs. (\ref{rn2}) and (\ref{rn1}), the moments $a_n$'s can be subsequently calculated recursively.
The first few values are
\begin{equation}
a_0=1, \quad\quad a_1=\frac{3}{4}\sqrt{\frac{\pi}{2}}, \quad\quad a_2= \frac{59}{60}, \quad\quad 
a_3=\frac{465}{512}\sqrt{\frac{\pi}{2}}, \quad \ldots
\label{memom1}
\end{equation}
We will use these results later in the context of the calculation of the moments
of the maximal height fluctuations in interfaces with free boundary conditions. 

\section{Fluctuating Edwards-Wilkinson Interface}

In this section, we will discuss some general properties of the dynamics and the 
stationary state of a fluctuating $(1+1)$-dimensional interface. 
We consider here the simplest model of a fluctuating one dimensional interface characterised
by a single valued height function $H(x,t)$ which is growing on a linear substrate of
size $L$ according to the celebrated Edwards-Wilkinson equation\cite{Edwards}
\begin{equation}
{{\partial H(x,t)}\over {\partial t}}= {{\partial^2 H(x,t)}\over {\partial x^2}}+\eta(x,t),
\label{ew1}
\end{equation}
where $\eta(x,t)$ is a Gaussian white noise with zero mean and a correlator, $\langle
\eta(x,t)\eta(x't')\rangle= 2 \delta(x-x')\delta(t-t')$. The equation (\ref{ew1}) has a soft
(zero wave vector) mode since the spatially averaged height ${\overline {H(x,t)}}= \int_0^{L}
H(x,t)dx/L$ diffuses with time (typical value growing as $\sqrt{t/L}$) even in a finite system
of size $L$. Thus, the joint height distribution $P[\{H\},t]$ of the system will never
reach a time independent stationary state. Hence, it is useful to subtract this zero mode from the 
height and define the relative
height, $h(x,t)= H(x,t)-{\overline {H(x,t)}}$, whose joint distribution $P[\{h\},t]$ will
eventually reach a stationary state in the long time limit in a finite system. 
Note that the average height ${\overline {H(x,t)}}$ is over a given sample, and not
the statistical average over the noise $\eta$ in Eq. (\ref{ew1}).
So, from now on,
we will always consider the relative height $h(x,t)$ as our basic variables. Note that, 
by definition,
\begin{equation}
\int_0^{L} h(x,t) dx =0.
\label{k0}
\end{equation}
We will see later that this constraint of zero total area under the relative height $h$ plays an
important role in determining
the maximal relative height (MRH) distribution. All the other nonzero modes of $h$ evolve 
identically as those of the actual height $H$.
 
We will see shortly that there are two regimes of the evolution of the interface in a finite system.
Let us assume that we start from a flat initial configuration where the heights at different points
on the substrate are completely uncorrelated from each other.  As time $t$ grows, the heights
at different points get correlated and the system is characterized by a single 
growing correlation length $\xi(t)\sim t^{1/z}$, where $z$ is the dynamical exponent 
($z=2$ for the EW interface). As long as 
this growing correlation length is much smaller than the system size, $\xi(t) \sim t^{1/z}<<L$,
the interface is in the {\em growing regime}. In this regime,
the system does not feel the boundary
 and hence is insensitive to the boundary conditions. When the correlation length becomes 
comparable or exceeds the system size, $\xi(t)\sim t^{1/z} \ge L$, the system crosses over
to a {\em stationary regime} where the joint distribution of relative heights $P[\{h\},t]$
becomes time independent. In this regime, the system is strongly sensitive to the boundary 
conditions. The crossover time between the two regimes scales as $t_c\sim L^z$.   

The two subsections below deal with two different boundary conditions, respectively the
periodic and the free boundary condition. Our goal in these subsections
would be to compute the height-height correlation function as well as the stationary joint 
distribution of the relative heights
$P[\{h\}]$ for the two boundary conditions. The form of these joint distributions
will be used later in Section IV for the calculation of the MRH distribution in the stationary state.

\subsection{ Periodic Boundary Conditions}

We first consider the periodic boundary condition, $H(0,t)=H(L,t)$. 
The relative height $h(x,t)= H(x,t)- \int_0^{L} H(x,t)dx/L$ satisfies the same boundary condition, $h(0,t)=h(L,t)$
and also satisfies the same evolution equation (\ref{ew1}) as the actual height $H(x,t)$.
Since the relative height $h(x,t)$ is a periodic function, one can decompose it
into a Fourier series, $h(x,t)= \sum_{m=-\infty}^{\infty} {\tilde
h}(m,t)e^{2\pi i mx/L}$. Substituting this in Eq. (\ref{ew1}), one finds that
different nonzero Fourier modes decouple from each other and one can easily calculate
any correlation function. For example, the height-height correlation function is given by 
the exact expression,
\begin{equation}
C(x=|x_1-x_2|, t, L)=\langle h(x_1,t)h(x_2,t)\rangle= \frac{L}{4\pi^2}\sum_{m\ne 0} 
\frac{1}{m^2}\left[1-e^{-8\pi^2 m^2t/L^2}\right] e^{2\pi i m(x_1-x_2)/L},
\label{ewc1}
\end{equation}
where the sum ranges from $m=-\infty$ to $m=\infty$ excluding the $m=0$ term. 

In the {\em growing regime} when
$t<<L^2$, we can make a change of variable $k=2\pi m \sqrt{t}/L$. Since $t<<L^2$, $dk=2\pi \sqrt{t}/L $ is 
small
and hence $k$ can be considered as a continuous variable. One 
can then replace the sum in eq. (\ref{ewc1}) by an integral over the $k$ space. This integral can be 
easily 
performed and one gets,
\begin{equation}
C(x,t) = \sqrt{t}\, G\left({x}/{\sqrt t}\right),
\label{ewc2}
\end{equation}
with the scaling function, $G(y)= \sqrt{\frac{2}{\pi}} e^{-y^2/8} - 
\frac{1}{2} \,y\, {\rm erfc}\left(y/\sqrt{8}\right)$,
where ${\rm erfc}(x)= \frac{2}{\sqrt \pi}\int_x^{\infty} e^{-u^2}du$ is the 
complementary 
Error function. The scaling function $G(y)$ starts at $G(0)=\sqrt{2/{\pi}}$ and decays 
for large $y$ as $G(y) \approx \frac{4\sqrt 2}{\sqrt \pi y^2}e^{-y^2/8}$.  The onsite 
variance grows as , $\langle h^2(0)\rangle = C(0,t)\sim t^{1/2}$ indicating that the typical relative height
in this regime grows as a power law, $h\sim t^{\beta}$ where the growth exponent $\beta=1/4$ for the
$(1+1)$-dimensional EW interface. 
The height-height correlation between two sites, at a finite time $t<< L^2$ and separated by a large 
distance $1<< x<< L$, decays as
$C(x,t)\approx \frac{4{\sqrt 2} t^{3/2} }{\sqrt \pi x^2} e^{-x^2/8t}$. Thus, in this growing 
regime, the correlations decay over a length scale $\xi(t)\sim t^{1/2}<< L $ which is much smaller
than the system size $L$.
 
In the opposite {\em stationary regime} when $t>> L^2$, one can drop the exponential factor in Eq. 
(\ref{ewc1}). The correlation function becomes time independent. Summing the resulting 
series in Eq. (\ref{ewc1}) one gets,
\begin{equation}
C(x,L) = \frac{L}{12}\left[1- \frac{6x}{L}\left(1-\frac{x}{L}\right)\right].
\label{ewc3}
\end{equation}
Thus, the onsite variance $\langle h^2(0)\rangle = C(0,L)= L/12$. It is evident from Eq. 
(\ref{ewc3}) that in the {\em stationary regime}, the heights at different space points are 
{\em strongly} correlated. In Section V, we will calculate exactly the distribution of
the maximum of these heights in the stationary regime. Thus, our result will provide an
exact solution of the distribution of the extremum of a set of strongly correlated
random variables.

We are now ready to write down the joint distribution of heights $P\left[\{h\}\right]$ in the stationary state. 
For simplicity, let us first consider the single site stationary height distribution $P(h,t\to \infty)$. 
Because of the linearity of Eq. (\ref{ew1}) the stationary measure on $h$ is Gaussian.
Using $\langle h^2(0)\rangle = C(0,L)= L/12$, the stationary single site height distribution is 
Gaussian,
\begin{equation}
P_{\rm st}(h)= \sqrt{\frac{6}{\pi L}}e^{-6h^2/L}.
\label{ewsd1}
\end{equation}
Moreover, from Eq. (\ref{ewc1}), one can easily show that the slope-slope correlation function,
$\langle \partial_x h \partial_{x'} h\rangle \to \delta(x-x')-1/L$
in the stationary state. The local slopes $\partial_x h$ are thus uncorrelated
in the stationary state, except for the overall constraint due to the periodic boundary condition, $\int_0^L dx\,
\partial_x h =0$, that gives rise to the residual $1/L$ term. 
Collecting these facts together and remembering the constraint in Eq. (\ref{k0}), one readily writes down
the stationary joint distribution of heights (a multivariate Gaussian distribution),
\begin{equation}
P\left[\{h\}\right]= A_L \, e^{-{1\over {2}}\int_0^{L}d\tau (\partial_{\tau} h)^2}\,
\delta\left[h(0)-h(L)\right]\, \delta\left[ \int_0^{L} h(\tau)d\tau\right],
\label{mpbc}
\end{equation}
where $A_L$ is a normalization
constant and the two delta functions take care respectively of the periodic boundary condition
$h(0)=h(L)$ and the zero area constraint in Eq. (\ref{k0}). Note that in Eq. (\ref{mpbc}), $\tau$
refers to the space (and not time) and varies over the interval $0\le \tau\le L$. In fact, from now on
we will use $\tau$ to denote the space in order to make correspondence with the results derived in Sec. II. 

Before proceeding to calculate the normalization constant $A_L$, we make one important remark here.
The exponential factor inside the probability measure in Eq. (\ref{mpbc}) indicates that the
stationary paths are locally Brownian, i.e., evolve in space as, $dh(\tau)/d\tau= \xi(\tau)$, where $\xi(\tau)$ is a 
Gaussian white noise with zero mean and a correlator,
$\langle \xi(\tau)\xi(\tau')\rangle = \delta(\tau-\tau')$. 
For the periodic boundary condition, the stationary
path in space is, in fact, a Brownian bridge. 
This stationary Brownian measure also follows rather obviously even from the basic EW equation in Eq. (\ref{ew1}) whose 
equilibrium Gibbs-Bolzmann
state precisely corresponds to the exponential factor in Eq. (\ref{mpbc}). 
However, one has to be a bit more careful. The multiplicative factor representing the zero mode
constraint, $\delta\left[ \int_0^{L}
h(\tau)d\tau\right]$, in Eq. (\ref{mpbc}) indicates that {\em only those Brownian bridges should be sampled 
which enclose a total zero area over the interval} $[0,L]$.
Normally, in writing down the stationary measure
for the relative heights, one ignores this multiplicative delta function factor 
simply due to the fact that in most of the calculations on the 
interfaces, such as in the calculation of the distribution of the stationary width\cite{Width}, this constraint
does not play any important role. However, as we will see later in Section IV, this constraint does indeed play a
major role in determining the {\em maximal} height distribution
in the stationary state. Hence, it is worth
being careful in writing the full exact stationary joint distribution of the relative heights keeping all
the factors explicitly as done in Eq. (\ref{mpbc}). 

{\bf The normalization constant} $A_L$ : To determine the normalization constant, we note that if
one integrates over the heights at all the intermediate points in Eq. (\ref{mpbc}) but keeping the heights at the two
ends fixed at, say, $h(0)=h(L)=u$, one should recover the stationary single point height distribution $P_{\rm st}(h_0=u)$ 
in Eq. 
(\ref{ewsd1}), .i.e.,
\begin{equation}
A_L\, \int_{h(0)=u}^{h(L)=u} {\cal D} h(\tau)\,e^{-{1\over {2}}\int_0^{L}d\tau (\partial_{\tau} h)^2}\,
\delta\left[ \int_0^{L} h(\tau)d\tau\right]
= \sqrt{\frac{6}{\pi L}}e^{-6u^2/L}.
\label{ewn1}
\end{equation}

To evaluate the path integral on the left hand side of Eq. (\ref{ewn1}), we use a 
simple and 
elegant method by identifying the path integral as the propagator of a random acceleration process that has 
been well studied\cite{Burkhardt}.
To make the connection between the two problems, we first note that in the stationary state,  
the height $h(\tau)$ locally evolves in space as a Brownian motion,
$dh(\tau)/d\tau= \xi(\tau)$ where $\xi(\tau)$ is a Gaussian white noise with zero mean and a correlator,
$\langle \xi(\tau)\xi(\tau')\rangle = \delta(\tau-\tau')$. But, as mentioned earlier, this by itself is not enough since it 
does not take into 
account the important delta function constraint $\delta\left[ \int_0^{L} h(\tau)d\tau\right]$. To 
incorporate this special condition, let us 
define a new variable, $X(\tau)= \int_0^{\tau} h(\tau)d\tau$. The variable $X(\tau)$ then evolves as,
$d^2X/d\tau^2 = dh/d\tau= \xi(\tau)$ and hence can be identified as the position of a particle which
is randomly accelerated. Thus, to take into account the
full probability measure in Eq. (\ref{mpbc}), one has to consider the joint evolution, in space, of both
the variables $[X(\tau), h(\tau)]$ where $X(\tau)$ is the position and the $h(\tau)= dX/d\tau $ is the velocity
of the random accelerator. The joint propagator of this random acceleration process
$G[X,h,T|X_0,h_0,0]$, i.e., the probability that the process reaches its final value $[X,h]$ in time $T$ 
starting from its initial value $[X_0,h_0]$ at time $0$, can be easily calculated and is well 
known\cite{Burkhardt}
\begin{equation}
G[X,h,T|X_0,h_0,0]= \frac{\sqrt{3}}{\pi T^2}\, \exp\left[-\frac{6}{T^3}\left(X-X_0-h_0 T\right)\left(X-X_0-h T\right)
-\frac{2}{T}{\left(h-h_0\right)}^2\right].
\label{pracc}
\end{equation}

The next step is to note that the path integral on the left hand side of Eq. (\ref{ewn1}) is just the propagator
of the random acceleration problem in Eq. (\ref{pracc}), going from its initial value $[X_0=0, h_0=u]$ to its final value 
$[X=0, h=u]$
in `time' $L$ and hence is given by,
\begin{equation}
\int_{h(0)=u}^{h(L)=u} {\cal D} h(\tau)\, e^{-{1\over {2}}\int_0^{L}d\tau (\partial_{\tau} h)^2}\,
\delta\left[ \int_0^{L} h(\tau)d\tau\right]= G[0,u,L|0,u,0]=\frac{\sqrt{3}}{\pi L^2}\,
\exp\left[-6u^2/L\right].
\label{ewn2}
\end{equation}
Substituting this result in Eq. (\ref{ewn1}) and cancelling out the exponential factors from both sides, one gets 
\begin{equation}
A_L = \sqrt{2\pi}\, L^{3/2}.
\label{pnc1}
\end{equation}

\subsection{ Free Boundary Conditions}

In this subsection, we consider the free boundary condition where the slopes of 
the interface vanish at the two ends, $\partial_x h =0$ at $x=0$ and $x=L$. This boundary condition arises naturally 
if one considers a spatially discretized version of the continuum EW equation, 
\begin{equation}
{{dh(i,t)}\over {dt}}= h(i+1,t)+h(i-1,t)-2 h(i,t) + \eta(i,t),
\label{fbc1}
\end{equation}
where $i=1,2\dots,L$ and $h(i,t)$ can be interpreted as the displacement of the $i$-th bead of a polymer chain where the 
beads are connected via harmonic springs, such as in the Rouse model\cite{Rouse}. If the chain forms a cycle, one has the 
periodic boundary condition. On the other hand, if the two ends of the chain are free, the beads at the two end points 
feel only one sided interactions, i.e.,
\begin{eqnarray}
{{dh(L,t)}\over {dt}}&=& h(L-1,t)-h(L,t) + \eta(L,t), \nonumber \\ 
{{dh(1,t)}\over {dt}} &=& h(2,t)-h(1,t) + \eta(0,t).
\label{fbc2}
\end{eqnarray}
The end point equations (\ref{fbc2}) will be consistent with Eq. (\ref{fbc1}) provided one incorporates the boundary
conditions, $h(L,t)=h(L+1,t)$ and $h(1,t)=h(0,t)$. In the continuum space, this boundary condition is equivalent
to having $\partial_x h =0$ at $x=0$ and $x=L$.

The zero slope conditions at the two end points indicate that one should decompose the height variable into
a cosine series, $h(x,t)= \sum_{m=1}^{\infty} {\tilde
h}(m,t)\cos(m\pi x/L)$. As in the periodic case, the different modes evolve independently and one can easily calculate 
the height-height correlation function. We find,
\begin{equation}
C(x_1,x_2,t, L)= \langle h(x_1,t)h(x_2,t)\rangle= \frac{2L}{\pi^2}\sum_{m=1}^{\infty} 
\frac{1}{m^2}\left[1-e^{-2m^2\pi^2t/L^2}\right] \cos \left(\frac{m\pi x_1}{L}\right)\cos \left(\frac{m\pi x_2}{L}\right).
\label{fewc1}
\end{equation}
Note that, unlike the periodic case in Eq. (\ref{ewc1}), the correlation function $C(x_1,x_2,t,L)$ is not translationally 
invariant and depends on both co-ordinates $x_1$ and $x_2$. 

In the {\em growing regime} when $t<< L^2$, one can again replace the sum in Eq. (\ref{fewc1}) by an integral over
the $k=\pi m/L$ space. The resulting integral can be easily performed and one finds,
\begin{equation}
C(x_1, x_2, t) = \sqrt{t}\, \left[G\left( \frac{|x_1-x_2|}{\sqrt t}\right)+G\left( \frac{|x_1+x_2|}{\sqrt t}\right)\right],
\label{fewc2}
\end{equation}
with the same scaling function, $G(y)= \sqrt{\frac{2}{\pi}} e^{-y^2/8} - 
\frac{1}{2}\,y\, {\rm erfc}\left(y/\sqrt{8}\right)$,
as in the periodic case. Note that in the limit when both $x_1$ and $x_2$ are large (but both small compared to $L$) 
with the distance between them $x=|x_1-x_2|$ fixed, Eq. (\ref{fewc2}) reduces to the same correlation function as in 
the periodic case in Eq. (\ref{ewc2}). This is physically expected, since in the growing regime $t<<L^2$, the system is 
insensitive to the boundary conditions.

In the {\em stationary regime} when $t>> L^2$, the time independent correlation function can be obtained by dropping
the exponential factor in Eq. (\ref{fewc1}) and summing the resulting series. We get,
\begin{equation}
C(x_1,x_2,L)=L \left[\frac{1}{3} - \frac{1}{2L}\left(x_1+x_2+|x_1-x_2|\right) + 
\frac{1}{2L^2}\left(x_1^2+x_2^2\right)\right].
\label{fewc3}
\end{equation}
In particular, the onsite variance, $\langle h^2(x)\rangle = C(x,x,L)= 
L\left[\frac{1}{3}-\frac{x}{L}\left(1-\frac{x}{L}\right)\right]$ now depends on the site co-ordinate $x$ since the
translational invariance is lost with free boundary conditions. Note that, as in the periodic case, the heights are 
{\em strongly} correlated in the stationary state.

The stationary single point height distribution is Gaussian,
$P_{\rm st}\left(h(x)\right)=e^{-h^2(x)/{2C(x,x,L)}}/{\sqrt{2\pi C(x,x,L)}}$. 
In particular, the stationary height distribution at $x=0$ is given by
\begin{equation}
P_{\rm st}(h(0))=\sqrt{\frac{3}{2\pi L}}\, e^{-3h^2(0)/{2L}}.
\label{fewsd1}
\end{equation}
One can also calculate the slope-slope correlation function from Eq. (\ref{fewc1}) and one finds that for $t>>L^2$, 
$\langle \partial_x h \partial_{x'} h\rangle \to \delta(x-x')$, implying that the slopes get completely uncorrelated
in the stationary state. Once again, collecting these facts together and taking into account the constraint in Eq. 
(\ref{k0}), one finds that the 
stationary joint distribution of heights for the free boundary case, for fixed boundary heights $h(0)=u$ and 
$h(L)=v$, is given by
\begin{equation}
P\left[\{h\}, u,v\right]= B_L \, e^{-{1\over {2}}\int_0^{L}d\tau (\partial_{\tau} h)^2}\,
\delta\left[ \int_0^{L} h(\tau)d\tau\right]\, \delta\left[h(0)-u\right]\,\delta\left[h(L)-v\right]
\label{mfbc}
\end{equation}
where the normalization constant $B_L$ is yet to be determined.

{\bf The normalization constant} $B_L$ : The constant $B_L$ is calculated using the similar method as in the periodic case.
First, we integrate over the heights at all the intermediate points in Eq. (\ref{mfbc}) but keeping the heights at 
the two ends fixed, say $h(0)=u$ and $h(L)=v$. This gives us the joint stationary distribution of heights at the
two end points, $P_{\rm st}[u,v]$. If we now further integrate over $v$, we should get the single point 
height distribution $P_{\rm st}[h(0)=u]$ given in Eq. (\ref{fewsd1}). Thus we have,
\begin{equation}
B_L \int_{-\infty}^{\infty} dv \int_{h(0)=u}^{h(L)=v} {\cal D} h(\tau)\,e^{-{1\over {2}}\int_0^{L}d\tau (\partial_{\tau} 
h)^2}\,
\delta\left[ \int_0^{L} h(\tau)d\tau\right]=\sqrt{\frac{3}{2\pi L}}e^{-3u^2/{2L}}.
\label{fewn1}
\end{equation}
As in the periodic case, we now identify the path integral on the left hand side of Eq. (\ref{fewn1}) as the
propagator $G[0,v,L|0,u,0]$ of the random acceleration process in Eq. (\ref{pracc}) going from the initial
point $[0,u]$ to the final point $[0,v]$ in `time' L. This gives,
\begin{equation}
\int_{h(0)=u}^{h(L)=v} {\cal D} h(\tau)\,e^{-{1\over {2}}\int_0^{L}d\tau (\partial_{\tau}
h)^2}\,
\delta\left[ \int_0^{L} h(\tau)d\tau\right]=G[0,v,L|0,u,0]= \frac{\sqrt{3}}{\pi L^2}\, 
\exp\left[-\frac{6}{L}uv-\frac{2}{L}{\left(u-v\right)}^2\right].
\label{fewn2}
\end{equation}
We substitute this expression of the path integral on the left hand side of Eq. (\ref{fewn1}) and then 
integrate over $v$. Cancelling the exponential factors from both sides of Eq. (\ref{fewn1}) gives
\begin{equation}
B_L= L .
\label{fnc1}
\end{equation}

\section{Maximal Relative Height Distribution in the Stationary Regime: Edwards-Wilkinson Interface}

In this section, we derive the principal new results of this paper, namely the exact probability distribution
of the maximal relative height (MRH) in the stationary state of the Edwards-Wilkinson equation in a finite one 
dimensional system of 
size $L$, both for periodic and free boundary conditions. 

\subsection{ Periodic Boundary Conditions}

Consider the Edwards-Wilkinson model in Eq. (\ref{ew1}) in the stationary regime when $t>> L^2$. The 
exact form of the stationary 
joint distribution of the relative heights $P\left[\{h\}\right]$ is given in Eq. (\ref{mpbc}) with
the normalization constant, $A_L=\sqrt{2\pi}\, L^{3/2}$. We wish to compute the probability distribution 
of the MRH, ${\rm max}\left[\{h\}\right]$. Let us first define the cumulative distribution of the MRH, $F(h_m,L)= 
{\rm
Prob}\left[ {\rm max}\{h\}<h_m, L\right]$. The distribution of the MRH is simply the derivative,
$P(h_m,L)= {{\partial F(h_m,L)}/ {\partial h_m}}$. Clearly $F(h_m,L)$ is also the probability that the 
heights at all points in $[0,L]$ are less than $h_m$ and hence can be written using the measure in Eq. 
(\ref{mpbc}) as a path integral,
\begin{equation}
F(h_m, L)= A_L \int_{-\infty}^{h_m} d u\, \int_{h(0)=u}^{h(L)=u} {\cal D} h(\tau)\,
 e^{-{1\over {2}}\int_0^{L}d\tau (\partial_{\tau} h)^2}\, \delta\left[ \int_0^{L} h(\tau)d\tau\right] I(h_m,L),
\label{pcum1}
\end{equation}
where $I(h_m, L)= \prod_{\tau=0}^{L}\theta(h_m-h(\tau))$ is an indicator function
which is $1$ if all the heights are less than $h_m$ and zero otherwise.
Due to the periodic boundary condition, $h(0)=h(L)$, all
the paths inside the path integral propagate from their initial value $h(0)=u$
to their final value $h(L)=u$. Note also that one needs to integrate over the boundary value $h(0)=h(L)=u$
over the allowed range $-\infty\le u \le h_m$. The upper bound $u\le h_m$ follows from the fact that, by 
definition, $h_m$ is the maximum over all heights in $[0,L]$ including the two end points.

The next crucial step is to make a change of variables, $x(\tau)= h_m-h(\tau)$ and $u'=h_m-u$ in the path
integral in Eq. (\ref{pcum1}) which gives,
\begin{equation}
F(h_m, L)= A_L \int_{0}^{\infty} d u'\, \int_{x(0)=u'}^{x(L)=u'} {\cal D} x(\tau)\,
 e^{-{1\over {2}}\int_0^{L}d\tau (\partial_{\tau} x)^2}\,
\delta\left[ \int_0^{L} x(\tau)d\tau -A \right] I(L),
\label{pcum2}
\end{equation}
where $I(L)= \prod_{\tau=0}^{L} \theta(x(\tau))$ and $A=h_m L$. Physically, this change of variable
just means that we are now measuring the interface height relative to the maximum and denoting it by
$-x(\tau)$. The maximal height now corresponds to the level $x=0$. Note that
$h_m$ appears only through the quantity $A$ in the delta function, and hence
$F(h_m, L) = {\cal F}(A, L)$. In the subsequent
analysis, we will keep a general $A$ in Eq. (\ref{pcum2}) and will substitute
$A=h_m L$ only in the final formula. 

We already start seeing the formal similarity between
Eq. (\ref{pcum2}) and Eq. (\ref{exarea2}) in Section II for the distribution of area
under a Brownian excursion. To be more precise,  
we take the Laplace transform with respect to $A$ in Eq. (\ref{pcum2}) which gives,
\begin{equation}
\int_0^{\infty} {\cal F}(A,L)e^{-\lambda A}dA = A_L \int_{0}^{\infty} du' <u'|e^{-{\hat H}_1 L}|u'>= 
A_L \, {\rm Tr}\left[e^{-{\hat H}_1 L} \right],
\label{trace1}
\end{equation}
where ${\rm Tr}$ is the trace and ${\hat H}_1$ is precisely the same Hamiltonian that appeared in Section II 
namely,
${\hat H}_1\equiv -\frac{1}{2} \frac{d^2}{dx^2} +V_1(x)$
where $V_1(x)$ is a triangular potential: $V_1(x)= \lambda x$ for $x>0$ and $V_1(x)=\infty$ for $x\le
0$. The Hamiltonian ${\hat H}_1$ has only discrete eigenvalues that were obtained in Section II,
$E_k= \alpha_k \lambda^{2/3}
2^{-1/3}$ for $k=1,2,\ldots$, where $\alpha_k$'s are the magnitude of the zeros of $Ai(z)$
on the negative real axis. The trace in Eq. (\ref{trace1}) can be easily evaluated knowing these eigenvalues
and one gets using $A_L= \sqrt{2\pi}\, L^{3/2}$,
\begin{equation}
\int_0^{\infty} {\cal F}(A,L)e^{-\lambda A}dA = \sqrt{2\pi}\, L^{3/2} \sum_{k=1}^{\infty} e^{-\alpha_k 
\lambda^{2/3} 2^{-1/3} L}. 
\label{plt1}
\end{equation}
Next we invert the Laplace transform in Eq. (\ref{plt1}) using Bromwitch formula
and substituting $A=h_m L$ we get
the cumulative MRH distribution,
\begin{equation}
F(h_m,L)= \sqrt{2\pi} L^{3/2} \int_{\lambda_0-i\infty}^{\lambda_0+i \infty} {{d\lambda}\over {2\pi i}} e^{\lambda
h_m L} \sum_{k=1}^{\infty} e^{-\alpha_k \lambda^{2/3} 2^{-1/3} L},
\label{ilt1}
\end{equation}
where the integration is along any imaginary axis whose real part $\lambda_0$ must be to the
right of all singularities of the integrand.
Taking derivative with respect to $h_m$ in Eq. (\ref{ilt1}) and making a change
of variable, $\lambda= sL^{-3/2}$, we arrive
at our main result,
$P(h_m,L)= L^{-1/2} f\left(h_m L^{-1/2}\right)$ for all $L$, where the Laplace transform of
$f(x)$ is given by
\begin{equation}
\int_0^{\infty} f(x) e^{-sx} dx = s\sqrt{2\pi} \sum_{k=1}^{\infty} e^{-\alpha_k s^{2/3}2^{-1/3}}.
\label{lt1}
\end{equation}
Comparing with Eq. (\ref{exarea5}) in Section II, we identify the scaling function $f(x)$ as the Airy 
distribution function that characterizes the probability distribution of the area $a$ under a 
Brownian excursion over a unit interval.

Thus, the main result of this subsection has been to establish an equivalence in law between two rather different 
objects: (i) the scaled maximal relative height $h_m/\sqrt L$ in the stationary state of the EW equation 
(\ref{ew1}) over a
linear substrate of size $L$ and (ii) the area $a$ under a Brownian excursion over
the unit time interval $[0,1]$,
\begin{equation}
\frac{h_m}{\sqrt L}  \equiv a,
\label{equiv1}
\end{equation}
where $\equiv$ means that the object on the left hand side has the same probability distribution
as the object on its right hand side.

{\bf Moments of the MRH:} Using the scaling form, $P(h_m, L)= L^{-1/2} f\left(h_m L^{-1/2}\right)$ for all $L$,
we get $\langle h_m^{n}\rangle = M_n L^{n/2}$ where $M_n=\int_0^{\infty} f(x) x^n dx$ are the moments of the Airy 
distribution function that were computed in Eq. (\ref{exmoment2}) of Section II. 
In the context of the MRH, only the
second moment $\langle h_m^{2}\rangle=5L/12$ was computed previously in \cite{RCPS} by using a different
method. Here, we can compute all the moments of MRH recursively.

{\bf Asymptotic tails of the MRH distribution:} In order to determine the asymptotic behaviors of the
MRH distribution, 
$P(h_m, L)= L^{-1/2} f\left(h_m L^{-1/2}\right)$, we need to know
the tails of the function $f(x)$.  As 
mentioned in the introduction, Takacs was able to formally 
invert the Laplace transform in Eq. (\ref{airy1}) and obtained an expression of $f(x)$ in terms of confluent
hypergeometric functions as given in Eq. (\ref{fx1}).  
It is easy to obtain the small $x$ behavior of $f(x)$ from Eq.
(\ref{fx1}), since only the $k=1$ term dominates the sum near $x\to 0$. Using $U(a,b, z)\sim z^{-a}$ for
large $z$\cite{AS}, we get as $x\to 0$,
\begin{equation}
f(x)\to {8 \over {81}} \alpha_1^{9/2} x^{-5}\,\exp\left[-{{2\alpha_1^3}\over {27 x^2}}\right],
\label{es1}
\end{equation}
where $\alpha_1=2.3381074\dots$. In the context of the MRH distribution,
this essential singular tail of the scaling function near $x\to 0$ was conjectured in \cite{RCPS} based
on numerics, though the exact form was missing.
The asymptotic behavior at large $x$ is more
tricky to derive\cite{CSY} from
Eq. (\ref{fx1}). However, it is possible to guess the leading behavior of $f(x)$ for large $x$ by examining 
the behavior of moments $M_n$. Using the recursion relations for the moments $M_n$ in Section II (Eq. 
(\ref{exmoment1})), one finds that for large $n$, $M_n \sim (n/12e)^{n/2}$. Substituting
an anticipated large $x$ decay of the form, $f(x)\sim \exp[- a x^b]$ in
$M_n=\int_0^{\infty} f(x) x^{n}dx$, we get $M_n\sim (n/{abe})^{n/b}$ for
large $n$. Comparing this with the exact large $n$ behavior of $M_n$ we get $a=6$ and $b=2$,
indicating a Gaussian tail as $x\to \infty$,
\begin{equation}
f(x) \sim e^{-6 x^2}.
\label{pg1}
\end{equation}
However, we are not able to determine the amplitude (or the next subleading correction) of this Gaussian tail.

\begin{figure}[htbp]
\epsfxsize=8cm
\centerline{\epsfbox{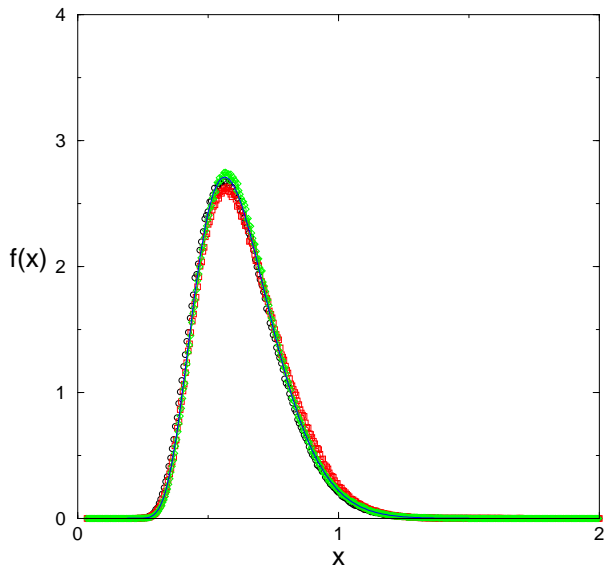}}
\caption{The scaling function $f(x)$ associated with the MRH distribution , $P(h_m, L)= L^{-1/2}f(h_m/\sqrt{L})$
for the EW equation with periodic boundary condition. The numerical curves (shown by symbols) are obtained by collapsing 
the data obtained
from the numerical integration of Eq. (\ref{dew1}) for three system sizes $L=256$ (circles), $L=384$ (squares), 
and $L=512$ (diamonds). Also 
plotted is the Mathematica generated analytical scaling function in Eq. (\ref{fx1}) as shown by the solid line.}
\label{fig:pbcewd}
\end{figure}
{\bf Comparison with numerical simulations:} Finally, we would like to compare our exact result for the MRH
distribution, $P(h_m,L)= L^{-1/2} f\left(h_m
L^{-1/2}\right)$ with $f(x)$ given in Eq. (\ref{fx1}), to the results of numerical simulations.
We numerically integrated the space-time discretized form of Eq. (\ref{ew1}),
\begin{equation} 
H(i, t+\Delta t)-H(i,t)= \Delta t \left[ H(i+1, t)+ H(i-1,t) -2 H(i,t) \right]+ \eta_i(t) \sqrt{2\Delta t},
\label{dew1}
\end{equation}
where $\eta_i(t)$'s are independent and identically distributed random variables for each $i$ and $t$ and
each drawn from a Gaussian distribution with zero mean and unit variance. 
We used periodic boundary condition, i.e., $H(0,t)=H(L,t)$ and $H(L+1,t)=H(1,t)$.
We chose the time step $\Delta t=0.01$, and we checked that the results were stable.
We used different system sizes $L=256$, $L=384$, and $L=512$.
We first evolved Eq. (\ref{dew1}) for a long time (typically $2\times 10^6$ Monte Carlo time steps) and ensured 
that the system has 
reached the stationary state. After that, the system was further evolved and the data for the 
histogram of the MRH was 
sampled at times which are far apart from each other to avoid correlations between them. 
The number of samples
used were typically $2\times 10^6$.
In Fig. 4, we have plotted the 
numerical scaling function generated by collapsing the data for the three system sizes and also plotted the
Airy distribution function $f(x)$ by evaluating the sum in Eq. (\ref{fx1}) using the Mathematica. The agreement
between the analytical and the numerical scaling function is very good.

\subsection{ Free Boundary Conditions}  

In this subsection we compute the stationary MRH distribution of the EW equation (\ref{ew1})
with free boundary conditions, $\partial_x h=0$ at $x=0$ and $x=L$. The stationary joint distribution of relative 
heights $P\left[\{h\},u,v \right]$ (with boundary heights fixed at $h(0)=u$ and $h(L)=v$) is given in Eq. (\ref{mfbc}) with 
$B_L=L$. As in the periodic case, we define the
cumulative MRH distribution, $F(h_m,L)={\rm Prob}\,\left[ {\rm max}\{h\}<h_m, L\right]$. Since $F(h_m,L)$ is also 
the probability that the
heights at all points in $[0,L]$ are less than $h_m$, one can use Eq. (\ref{mfbc}) to write
\begin{equation}
F(h_m, L)= L \int_{-\infty}^{h_m} du\, \int_{-\infty}^{h_m} dv \int_{h(0)=u}^{h(L)=v} {\cal D} h(\tau)\,
 e^{-{1\over {2}}\int_0^{L}d\tau (\partial_{\tau} h)^2}\, \delta\left[ \int_0^{L} h(\tau)d\tau\right] I(h_m,L)
\label{fcum1}
\end{equation}
where $I(h_m,L)= \prod_{\tau=0}^{L}\theta(h_m-h(\tau))$, as before, is an indicator function
which is $1$ if all the heights are less than $h_m$ and zero otherwise. Note that the boundary heights
$u$ and $v$ are integrated over their common allowed range $[-\infty, h_m]$. 

The next step is to make the change of variables, $x(\tau)=h_m-h(\tau)$, $u'=h_m-u$ and $v'=h_m-v$ where
$-x(\tau)$ now denotes the interface height measured relative to the maximal height. The maximal height now corresponds to 
the level $x=0$. With these change of variables, Eq. (\ref{fcum1}) now reads
\begin{equation}
F(h_m, L)= L \int_{0}^{\infty} du'\, \int_{0}^{\infty} dv' \int_{x(0)=u'}^{x(L)=v'} {\cal D} x(\tau)\,
 e^{-{1\over {2}}\int_0^{L}d\tau (\partial_{\tau} x)^2}\,
\delta\left[ \int_0^{L} x(\tau)d\tau -A \right] I(L)
\label{fcum2}
\end{equation}
where $I(L)= \prod_{\tau=0}^{L} \theta(x(\tau))$ and $A=h_m L$. The only dependence on $h_m$ in Eq. 
(\ref{fcum2}) appears through the quantity $A=h_m L$. Hence, $F(h_m, L) = {\cal F}(A, L)$. Taking Laplace transform with 
respect to $A$ in Eq. (\ref{fcum2}) gives,
\begin{equation}  
\int_0^{\infty} {\cal F}(A,L)e^{-\lambda A}dA = L \int_0^{\infty} du' \int_0^{\infty}dv' <u'|e^{-{\hat H}_1 L}|v'>,
\label{fcum3}
\end{equation}
where the Hamiltonian ${\hat H}_1$ is the same as in the periodic case. Eq. (\ref{fcum3}) for the free boundary case 
should be compared to the
corresponding result for the periodic case in Eq. (\ref{trace1}) where the integral over the common end points 
became a trace due to the periodic boundary condition. In the present case, we do not have a trace and evaluating
the integral in Eq. (\ref{fcum3}) would require the knowledge of both the eigenfunctions and the eigenvalues of
${\hat H}_1$. Fortunately, the eigenfunctions can be determined exactly in terms of the Airy function as explained in 
Section II. Expanding the propagator $<u'|e^{-{\hat H}_1 L}|v'>$ in Eq. (\ref{fcum3}) in the energy eigenbasis of ${\hat 
H}_1$ and using the exact eigenvalues $E_k= \alpha_k \lambda^{2/3}
2^{-1/3}$ for $k=1,2,\ldots$ ($\alpha_k$'s being the magnitude of the zeros of $Ai(z)$ on the negative real axis) we get,
\begin{equation}
\int_0^{\infty} {\cal F}(A,L)e^{-\lambda A}dA = L \sum_{k=1}^{\infty}e^{-\alpha_k \lambda^{2/3} 2^{-1/3} L} \int_0^{\infty} 
\psi_k(u')du' \int_0^{\infty} \psi_k (v') dv'.
\label{fcum4}
\end{equation}
Substituting the exact form of the eigenfunctions $\psi_k(x)$ from Eq. (\ref{exeigen1}) and simplifying, we get
\begin{equation}
\int_0^{\infty} {\cal F}(A,L)e^{-\lambda A}dA = \frac{L}{(2\lambda)^{1/3}}\sum_{k=1}^{\infty} C(\alpha_k) \, e^{-\alpha_k 
\lambda^{2/3} 
2^{-1/3} L}.
\label{fcum5}
\end{equation}
The constant $C(\alpha_k)$ is given by the expression,
\begin{equation}
C(\alpha_k)= \frac{ {\left[\int_{-\alpha_k}^{\infty} Ai(z)dz\right]}^2}{\int_{-\alpha_k}^{\infty}Ai^2(z)dz}= \frac{ 
{\left[\int_{-\alpha_k}^{\infty} Ai(z)dz\right]}^2}{Ai'^2(-\alpha_k)},
\label{fcum6}
\end{equation} 
where we have again used the identity\cite{Albright}, $\int_{-\alpha_k}^{\infty}Ai^2(z)dz= [Ai'(-\alpha_k)]^2$. Note that
$C(\alpha_k)= B^2(\alpha_k)$, where $B(\alpha_k)$ is given 
in Eq. (\ref{balphak}). The constants $C(\alpha_k)$'s
can be determined numerically very precisely from Eq. (\ref{fcum6}) using the Mathematica. 
As examples, we have determined the first few values in Table I. 
Note that $C(\alpha_k)$'s oscillate as $k$ increases.

\begin{center}
\begin{tabular}{||l||l||l||}  \hline
 $k$ & $\alpha_k$ & $C(\alpha_k)$ \\ \hline
1 & $2.33810\dots $ & $3.30279\dots $  \\
2 & $4.08794\dots $ & $1.01282\dots $  \\
3 & $5.52055\dots $ & $1.78218\dots $  \\
4 & $6.78670\dots $ & $0.90519\dots $  \\
5 & $7.94413\dots $ & $1.39473\dots $ \\
6 & $9.02265\dots $ & $0.83181\dots $ \\
7 & $10.04017\dots $ & $1.19915\dots $  \\
8 & $11.00852\dots $ & $0.77846\dots $ \\
9 & $11.93601 \dots $ & $1.07584\dots $ \\
10 & $12.82877\dots $ & $ 0.73725\dots $ \\ \hline
\end{tabular}
\end{center}

\noindent Table 1. The second column shows the magnitude $\alpha_k$ of the 
$k$-th zero of the Airy function $Ai(z)$ on the negative real axis upto $k=10$, obtained using
the Mathematica. The third column shows the corresponding values of the constant $C(\alpha_k)$
in Eq. (\ref{fcum6}), again generated via the Mathematica.

We formally invert the Laplace transform in Eq. (\ref{fcum5}), take the derivative with respect 
to $h_m$ and then make a change of variable, $\lambda= sL^{-3/2}$. This leads us to the exact MRH distribution for the free 
boundary case, $P(h_m, L)= L^{-1/2}\, F\left(h_m\, L^{-1/2}\right)$ for all $L$, where the Laplace transform of the 
scaling 
function $F(x)$ is given by,
\begin{equation}
\int_0^{\infty} F(x) e^{-sx} dx= 2^{-1/3} s^{2/3} \sum_{k=1}^{\infty} C(\alpha_k)\,  e^{-\alpha_k s^{2/3} 2^{-1/3}},
\label{flt1}
\end{equation}
with $C(\alpha_k)$'s determined from Eq. (\ref{fcum6}). One can check from Eq. (\ref{flt1}) that the function $F(x)$ is 
normalized, $\int_0^{\infty} F(x)dx=1$.

In order to make contact with the area under a Brownian meander discussed in Section II-B, we define a new variable
$t=2^{-1/3} s^{2/3}$ and rewrite Eq. (\ref{flt1}) as,
\begin{equation}
\int_0^{\infty} F(x) e^{-\sqrt{2} t^{3/2} x} dx = E\left[e^{-\sqrt{2} t^{3/2} x}\right]= \sum_{k=1}^{\infty} C(\alpha_k) 
t 
e^{-\alpha_k t},
\label{flt2}
\end{equation}
where $E$ denotes the expectation value with respect to the distribution $F(x)$. Taking a further Laplace transform of 
Eq. (\ref{flt2}) with respect to $t$ we get
\begin{equation}
\int_0^{\infty} e^{-ut} E\left[e^{-\sqrt{2} t^{3/2} x} \right] dt = \sum_{k=1}^{\infty} \frac{C(\alpha_k)}{(u+\alpha_k)^2}
= {\frac{\left[ \int_u^{\infty} Ai(z)dz \right]}{Ai(u)}}^2.
\label{flt3}
\end{equation}
The second equality in Eq. (\ref{flt3}) is proved in Appendix-A. 
Comparing Eq. (\ref{flt3}) with Eq. (\ref{marea6}) of Section II-B, one finds that the former is just the 
square of the latter. Using the result in Eq. (\ref{marea6}) and the convolution theorem of Laplace transform, we can then 
invert the Laplace transform in Eq. (\ref{flt3}) to obtain the following result,
\begin{equation}
E\left[ e^{-\sqrt{2} t^{3/2} x}\right]=\int_0^t \frac{dt'}{\pi \sqrt{t'(t-t')}}\, E\left[e^{-\sqrt{2} t'^{3/2} 
a_1}\right]\,
E\left[e^{-\sqrt{2} (t-t')^{3/2} a_2}\right],
\label{flt4}
\end{equation}
where $a_1$ and $a_2$ are the areas under two independent Brownian meanders each over unit intervals. The result in
Eq. (\ref{flt4}) is valid for all $t$. In particular, putting $t=L$ and $x=h_m/\sqrt{L}$, we get
\begin{equation}
E\left[ e^{-\sqrt{2} L h_m}\right]= \int_0^L \frac{dt'}{\pi \sqrt{t'(L-t')}}\, E\left[e^{-\sqrt{2} t'^{3/2}
a_1}\right]\,
E\left[e^{-\sqrt{2} (L-t')^{3/2} a_2}\right],
\label{flt5}
\end{equation} 
which has a very nice and simple physical interpretation. Consider the stationary interface profile over
the interval $[0,L]$ after we have made the shift with respect to the maximum, $x(\tau)= h_m-h(\tau)$.
A typical picture is shown in Fig. 5. 
\begin{figure}[htbp]
\epsfxsize=8cm
\centerline{\epsfbox{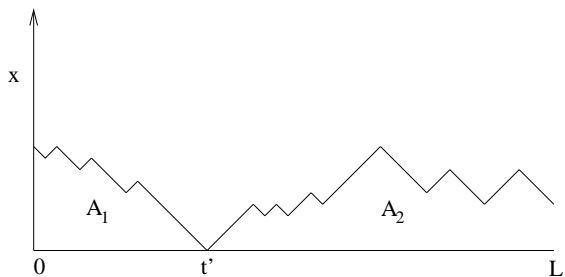}}
\caption{A stationary profile of the interface $x(\tau)=h_m-h(\tau)$ with the free boundary condition as seen from the
maximum $h_m$ located at $t'$. The point $t'$ divides the path into two statistically independent Brownian meanders
with areas $A_1$ and $A_2$ respectively. } 
\label{fig:conv}
\end{figure}

Let the location of the maximum be denoted by $t'$. Note that 
the location $0\le t'\le L$ of the maximum of a free Brownian motion over the interval $[0,L]$ is 
a random variable whose 
normalized probability distribution $P(t')$ is well known\cite{Feller},
\begin{equation}
P(t')= \frac{1}{\pi \sqrt{t'(L-t')}},
\label{mxb}
\end{equation}
which also describes, in the context of the interface with free boundary conditions, the distribution
of the position of the maximum of the interface.
It is clear from Fig. 5 that the maximum $h_m$ at $t'$ (or rather the minimum $x=0$ in Fig. 5) divides the path
into two Brownian meanders, one over the left interval $[0,t']$ and the other over the right interval
$[t', L]$ whose interval length is $(L-t')$. Since a Brownian motion is Markovian, these two meanders are 
statistically independent of 
each other. Thus, the areas under the two meanders, $A_1= a_1 t'^{3/2}$ and $A_2= a_2 (L-t')^{3/2}$ are also
statistically independent, except that their joining point $t'$ is distributed according to Eq. (\ref{mxb}).
Thus, Eq. (\ref{flt5}) has now a probabilistic interpretation,
\begin{equation}
h_m L \equiv \int_0^{L}\, \left[A_1 (t')+A_2 (L-t')\right] P(t')dt',
\label{equiv2}
\end{equation}
where $P(t')$ is given in Eq. (\ref{mxb}).
The random variable  $h_m L$ has therefore the same law as the sum of 
the two meander areas ( with the location of their joining point $t'$
integrated over with the measure $P(t')$ in Eq. (\ref{mxb})).
This result in Eq. (\ref{equiv2}) for 
the free boundary condition is thus
more involved than its periodic counterpart in Eq. (\ref{equiv1}).  

{\bf Moments of the MRH:} Using the scaling form, $P(h_m, L)= L^{-1/2}\,F(h_m\, L^{-1/2})$ for all $L$, we get
$\langle h_m^n\rangle =\mu_n L^{n/2}$, where $\mu_n= \int_0^{\infty} F(x) x^n dx$ are the moments of the function
$F(x)$ defined in Eq. (\ref{flt1}). The computation of $\mu_n$'s from Eq. (\ref{flt1}) is nontrivial. To make progress,
we make use of Eq. (\ref{flt4}). We note that the right hand side of Eq. (\ref{flt4}) involve the areas $a_1$ and $a_2$
of two independent
Brownian meanders over unit intervals. Now, the moments associated with the area of a Brownian meander over a unit 
interval $a_n= E[a^n]$ were computed by a recursive procedure in Section II-B and the results are given in Eq. 
(\ref{memom1}). By 
expanding the exponentials on the right hand side
of Eq. (\ref{flt4}) in power series and then performing the integral over $t'$ we get
\begin{equation}
E\left[ e^{-\sqrt{2} t^{3/2} x}\right]= \frac{1}{\pi}\sum_{n_1=0}^{\infty}\sum_{n_2=0}^{\infty} 
\frac{(-\sqrt{2})^{n_1+n_2}}{\Gamma(n_1+1)\Gamma(n_2+1)} B\left(\frac{3n_1+1}{2}, \frac{3n_2+1}{2}\right) a_{n_1} a_{n_2} 
t^{3(n_1+n_2)/2},
\label{fmom1}
\end{equation}
where $B(x,y)$ is the standard Beta function and the moments $a_n$'s of the area under a single meander over unit interval
were computed in Eq. (\ref{memom1}) in Section II-B. Next, expanding the left hand side of Eq. (\ref{fmom1}) in a power 
series and matching 
identical powers of $t$, we can express
the moment $\mu_n=\int_0^{\infty} F(x) x^n dx$ in terms of the $a_n$'s,
\begin{equation}
\mu_n = \frac{1}{\pi} \sum_{m=0}^{n} {n\choose m} B\left(\frac{3m+1}{2}, \frac{3(n-m)+1}{2}\right) a_m a_{n-m}.
\label{fmom2}
\end{equation}
Using the known values of $a_n$'s from Eq. (\ref{memom1}), 
one can then recursively determine the moments $\mu_n$ from Eq. (\ref{fmom2}). For example, the first few values are
\begin{equation}
\mu_0=1, \quad\quad \mu_1= \sqrt{\frac{2}{\pi}}, \quad\quad \mu_2=\frac{17}{24}, \quad\quad \mu_3= 
\frac{123}{140}\sqrt{\frac{2}{\pi}}, \quad \ldots   
\label{fmom3}
\end{equation}

{\bf Asymptotic tails of the MRH distribution:} To determine the asymptotic tails of the MRH distribution,
$P(h_m, L) = L^{-1/2} F\left(h_m L^{-1/2}\right)$, we need to know the asymptotic behavior of the scaling function $F(x)$
defined in Eq. (\ref{flt1}) in the limits $x\to 0$ and $x\to \infty$. Following similar method as Takacs in the periodic case 
in Eq. (\ref{fx1}), we were able to invert the Laplace transform in Eq. (\ref{flt1}) and express it in terms of the 
hypergeometric function. The derivation is presented in Appendix-B. We get,
\begin{equation}
F(x)= \frac{2^{-1/3}}{{\sqrt{3\pi}}\, x^{7/3}}\sum_{k=1}^{\infty} \alpha_k C(\alpha_k)\left[U\left(1/6,4/3,b_k/x^2\right)+ 2 
U\left(-5/6, 4/3, 
b_k/x^2\right)\right] e^{-b_k/x^2},
\label{frx1}
\end{equation} 
where $b_k= 2\alpha_k^3/{27}$, $C(\alpha_k)$ is defined in Eq. (\ref{fcum6}) and $U(a,b,z)$ is the confluent hypergeometric 
function. Thus, for the free boundary condition, the scaling function $F(x)$ is different from that of its periodic counterpart
namely the Airy distribution function in Eq. (\ref{fx1}). The function $F(x)$ appears to be new and hence it lacks a name. 
We call this function $F(x)$ as the $F$-Airy distribution function, where $F$ refers to the free boundary condition.  

In the limit $x\to 0$, only the $k=1$ term dominates the sum in Eq. (\ref{frx1}) and we get
\begin{equation}
F(x)\to {{2\sqrt {2}}\over {27\sqrt{\pi}}} C(\alpha_1) \alpha_1^{7/2} x^{-4}
\exp\left[-{{2\alpha_1^3}\over {27
x^2}}\right],
\label{es2}
\end{equation}
where $\alpha_1=2.33810\dots$ and $C(\alpha_1)=3.30278\dots$ from Table I, evaluated using the Mathematica. Thus the function 
$F(x)$ decays slightly 
faster as $x\to 0$ than its periodic counterpart in Eq. (\ref{es1}). Finding the asymptotic behavior of $F(x)$ as $x\to 
\infty$ turns out to be more tricky. However, as in the periodic case, the leading large $x$ behavior can be guessed
by analyzing the moment $\mu_n$ for large $n$. It is not difficult to analyze Eq. (\ref{fmom2}) for large $n$, knowing
the properties of $a_n$'s. We find that for large $n$, $\mu_n \sim [n/{2e}]^{n/2}$. Following exactly the same procedure 
as in the periodic case, we find that $F(x)$ has a leading Gaussian tail for large $x$,
\begin{equation} 
F(x)\sim e^{-3 x^2/2},
\label{fg1}
\end{equation}
that falls off less rapidly than the periodic case $f(x)\sim e^{-6x^2}$ in Eq. (\ref{pg1}).

\begin{figure}[htbp]
\epsfxsize=8cm
\centerline{\epsfbox{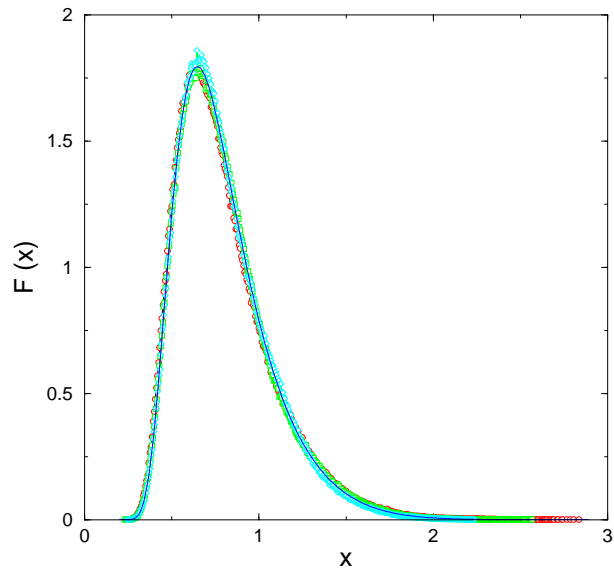}}
\caption{The scaling function $F(x)$ associated with the MRH distribution , $P(h_m, L)= L^{-1/2}F(h_m/\sqrt{L})$
for the EW equation with free boundary condition. The numerical curves (shown by symbols) are obtained by collapsing the data
obtained from the numerical integration of Eq. (\ref{dew1}) for three system sizes $L=256$ (circles), $L=384$ 
(squares), and $L=512$ (diamonds). Also
plotted is the Mathematica generated analytical scaling function in Eq. (\ref{frx1}) as shown by the solid line.}
\label{fig:fbcewd}
\end{figure}

{\bf Comparison with numerical simulations:} Our exact analytical result for the MRH distribution for the free boundary 
condition is: $P(h_m, L)= L^{-1/2} F\left(h_m L^{-1/2}\right)$ where the scaling function $F(x)$ is given in Eq. 
(\ref{frx1}). We have evaluated the sum in Eq. (\ref{frx1}) using the Mathematica. It turns out that the sum is rapidly 
convergent and it is sufficient to keep terms upto $k=10$ in the sum. 
We also numerically integrated the discretized Edwards-Wilkinson Eq. (\ref{dew1}) with
free boundary condition, i.e., $H(0,t)=H(1,t)$ and $H(L+1,t)=H(L,t)$ and calculated
the MRH distribution.
In Fig. 6, we compare the analytical scaling function
to the one obtained by numerical integration.
The number of steps used to reach the stationary state
as well as the number of sample averages were the same as in the case of the periodic boundary condition.
The numerical data shown in Fig. (\ref{fig:fbcewd}) were obtained by collapsing the
histograms for three system sizes $L=256$, $L=384$, and $L=512$.
The agreement is again very good.

\section{ Maximal Relative Height Distribution in the Stationary Regime: Kardar-Parisi-Zhang Interface}

In the previous section, we derived the exact distribution of the MRH in the stationary regime of the $(1+1)$-dimensional 
Edwards-Wilkinson interface evolving via Eq. (\ref{ew1}). A natural next step would be to extend our method to other
$(1+1)$-dimensional interfaces. In this section, we study the MRH distribution in the stationary regime of another
very well studied interface that evolves via the nonlinear KPZ equation\cite{KPZ},
\begin{equation}
{{\partial H(x,t)}\over {\partial t}}= {{\partial^2 H(x,t)}\over {\partial x^2}}+ \lambda {\left( {{\partial H(x,t)}\over 
{\partial x}} \right) }^2 + \eta(x,t),
\label{kpz1}
\end{equation}
where $\eta(x,t)$ is a Gaussian white noise with zero mean and a correlator, $\langle
\eta(x,t)\eta(x't')\rangle= 2 \delta(x-x')\delta(t-t')$. As before, we subtract the zero mode and focus on
the relative height, $h(x,t)= H(x,t)-{\overline {H(x,t)}}$ whose distribution reaches a stationary state
in the long time limit in a finite system.

It is well known that the dynamical exponent is $z=3/2$ for the $(1+1)$-dimensional KPZ equation\cite{HZ}. Thus, we expect that 
for times $t>> L^{3/2}$, the system will reach a stationary state. We wish to compute the distribution $P(h_m,L)$ of the
MRH, $h_m={\rm max}\left[\{h\}\right]$, in this stationary state. In order to compute this 
distribution, we 
need to know the stationary joint distribution of the relative heights $P\left[\{h\}\right]$ for the KPZ equation.
In an infinite system, it is well known\cite{HZ} that in $(1+1)$-dimensions, the joint distribution of heights
approaches a stationary state as $t\to \infty$, where $P\left[\{H\}\right]\propto \exp\left[-\frac{1}{2}\int (\partial_x H)^2 
dx\right]$. This is a somewhat surprising result since the dependence on the nonlinear term completely drops out of the 
stationary measure. Indeed, this turns out to be a very special property only 
valid in $(1+1)$-dimensions. This result can be 
proved by writing down the full Fokker-Planck equation for the joint height distribution $P\left[\{H\}, t \right]$ 
and then showing explicitly that indeed $P\left[\{H\}\right]\propto \exp\left[-\frac{1}{2}\int (\partial_x H)^2
dx\right]$ is a stationary solution of the Fokker-Planck equation\cite{HZ}. In proving the last step, one neglects
various boundary terms that arise out of integration by parts\cite{HZ} which is
justified only in the limit of an infinite system. However, in a finite system, the boundary terms are usually
nonzero and it is no longer easy to find the stationary measure.

The only exception seems to be the periodic boundary case where, even in a finite system, the boundary terms 
can be shown to vanish. Hence in this case the joint distribution of the relative heights $P\left[\{h\}\right]$ for 
the KPZ equation 
has the same expression as in the EW case, and is given by Eq. (\ref{ewn2}) with the normalization constant
$A_L=\sqrt{2\pi}\, L^{3/2}$. Therefore, for the periodic boundary condition,  we expect that the stationary MRH 
distribution for the KPZ equation 
will also be identical to that of the EW case, i.e., $P(h_m,L)= L^{-1/2} f\left(h_m L^{-1/2}\right)$ where the scaling 
function $f(x)$ is the Airy distribution function in Eq. (\ref{fx1}). 

An interesting challenge is to verify this analytical prediction numerically by directly integrating the KPZ 
equation (\ref{kpz1}). The problem is to find an appropriate spatial discretization scheme that will correctly represent
the nonlinear term in the continuum equation (\ref{kpz1}). Most of the past studies have used the following natural choice or 
its simple variants\cite{AF,MKD,DDK},
\begin{equation}
H(i, t+\Delta t)-H(i,t)= \Delta t \left[ \left(H\left(i+1, t\right)+ H\left(i-1,t\right) -2 H\left(i,t\right)\right) + 
\frac{\lambda}{4}{\left(H\left(i+1,t\right)-H\left(i-1,t\right)\right)}^2 \right]+ \eta_i(t) \sqrt{2\Delta t},
\label{dkpz1}
\end{equation}
where $\eta_i(t)$'s are independent and identically distributed random variables for each $i$ and $t$ and
each drawn from a Gaussian distribution with zero mean and unit variance. However, it is well known\cite{DDK,NB} that 
certain properties of Eq. (\ref{dkpz1}) are rather unphysical and are fundamentally different from those of the continum 
equation (\ref{kpz1}).
In other words, the discrete equation (\ref{dkpz1}) fails to capture the correct continuum KPZ evolution\cite{LS1}. 
In this paper, we have instead used a discretization scheme proposed by Lam and Shin\cite{LS2} which circumvents the
problems mentioned above. According to this scheme, one uses the equation (\ref{dkpz1}) except that the 
nonlinear term in Eq. (\ref{dkpz1}) is replaced by the following expression\cite{LS2},
\begin{equation}
\frac{\lambda}{3}\left[ {\left( H\left(i+1,t\right)-H\left(i,t\right)\right)}^2 + \left( 
H\left(i+1,t\right)-H\left(i,t\right)\right)\left(H\left(i,t\right)-H\left(i-1,t\right)\right) +
{\left( H\left(i,t\right)-H\left(i-1,t\right) \right)}^2\right].
\label{ls1}
\end{equation}
The advantage with this scheme is that one can prove analytically that the Fokker-Planck equation associated with this
discrete model admits a stationary solution for the periodic boundary case\cite{LS2}, $P\left[\{H\}\right]\propto 
\exp\left[-\frac{1}{2}\sum_{i=1}^L {\left(H\left(i+1\right)-H\left(i\right)\right)}^2\right]$ which, in the contiuum limit, 
correctly reproduces the stationary measure of the continuum 
KPZ equation, i.e., $P\left[\{H\}\right]\propto \exp\left[-\frac{1}{2}\int_0^L (\partial_\tau H)^2 d\tau\right]$.
We note that a similar but slightly different scheme that also leads to the Gaussian stationary state
has been proposed earlier\cite{KS2}.

We evolved the discretized equation (\ref{dkpz1}) (but with the nonlinear term in Eq. (\ref{dkpz1}) replaced by the 
Lam-Shin nonlinear term in Eq. (\ref{ls1})) with periodic boundary condition and 
chose $\lambda=1$ and $\Delta t=0.01$. 
The steady state MRH distribution $P(h_m, L)$ was obtained for three different system sizes $L=256$, $L=384$ and
$L=512$. The three data sets were collapsed onto a single scaling plot, $P(h_m, L)= L^{-1/2}f\left( h_m L^{-1/2}\right)$.
The numerically obtained scaling function is then compared in Fig. 7 with the analytical scaling function $f(x)$
given in Eq. (\ref{fx1}). The agreement is excellent.
Thus, for the periodic boundary condition, the stationary MRH distribution
of the $(1+1)$-dimensional KPZ equation is also described by the Airy distribution function, as in the case of the EW 
equation. However, we expect that this superuniversality of the Airy distribution function holds only in $(1+1)$-dimensions, 
and not in higher dimensions. 
\begin{figure}[htbp]
\epsfxsize=8cm
\centerline{\epsfbox{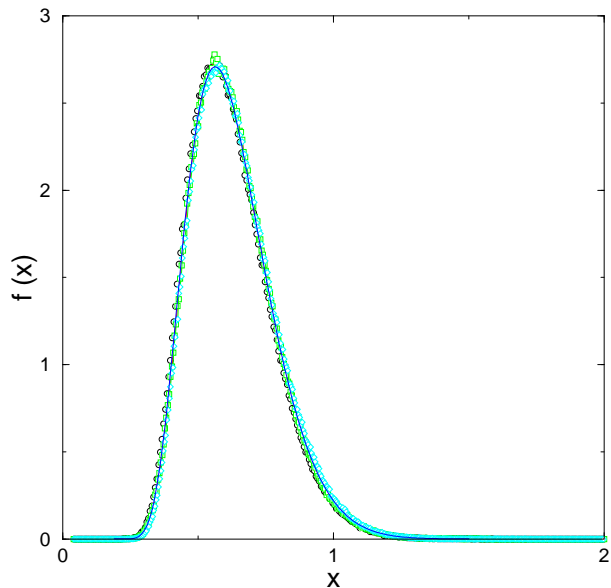}}
\caption{The scaling function $f(x)$ associated with the MRH distribution , $P(h_m, L)= L^{-1/2}f(h_m/\sqrt{L})$
for the KPZ equation with periodic boundary condition. The numerical curves (shown by symbols) are 
obtained by collapsing the data obtained
from the numerical integration of Eq. (\ref{dkpz1}) (but with the nonlinear term replaced by the Lam-Shin term from Eq. 
(\ref{ls1})) for three system 
sizes $L=256$ (circles), $L=384$ (squares), and $L=512$ (diamonds). Also
plotted is the Mathematica generated analytical scaling function in Eq. (\ref{fx1}) as shown by the solid line.}
\label{fig:pbckpz}
\end{figure}

We now turn to the case of the free boundary condition, i.e., with Neumann boundary
conditions $\partial_x h=0$ at the two ends $x=0$ and $x=L$. In the periodic case
$P[\{H\}]=\exp\left[-\frac{1}{2}\int (\partial_x H)^2\,dx\right]$ is explicitly
the stationary solution of the Fokker-Planck equation. However, for the free
boundary case, the same measure is no longer a stationary solution of the
Fokker-Planck equation. Thus, unlike in the periodic case where the Airy distribution
describing the MRH distribution for the Edwards-Wilkinson equation also describes
the MRH distribution of the KPZ equation, for the free boundary case there is
no apriori reason to believe that the F-Airy distribution for the Edwards-Wilkinson 
case will carry over to the KPZ case as well.  
In fact, as far as we know, the stationary measure for the
continuous KPZ equation with the Neumann boundary condition is still not known
explicitly. Thus, calculating analytically the corresponding MRH distribution
remains an interesting open problem. Let us also remark about the numerical
simulation with free boundary condition. As discussed before in this section, to simulate the continuous
KPZ equation one needs to use an appropriate discretization scheme that will
reproduce the correct continuum stationary measure. Arbitrary discretization scheme
can lead to spurious results. For the periodic 
boundary condition, 
such discretization schemes are available thanks to the fact that one knows the stationary state
explicitly. For the free boundary case where one does not know the stationary
state explicitly, it is apriori not clear what type of discretization schemes
one should use. 

In fact, to study the effects of different boundary conditions on the MRH distribution
in nonlinear KPZ type interfaces, it may be more appropriate and easier to study the discrete growth 
models that belong to the KPZ universality class, rather than the continuous KPZ equation 
itself. Several such models are known\cite{KS}, an example being the single-step
model\cite{Sstep}. This single-step model can further be mapped to the asymmetric
exclusion process (ASEP). A particle at a site $i$ in ASEP corresponds to a
downward slope of the interface height on one unit and a hole in ASEP corresponds
to an upward slope of the interface step of one unit. The height $h_i(t)$ at site $i$ in
the interface model 
is thus related to the occupancy $\tau_i$ in ASEP via the simple relation,
$h_{i+1}(t)-h_i(t)=1-2\tau_i(t)$ where $\tau_i=1$ (or $0$) if the site $i$ is occupied (or empty)
in the ASEP. Over the past few years there have been extensive studies of ASEP with open boundary 
conditions where a particle enters the lattice through its left end at rate $\alpha$
and leaves the lattice throught its right boundary at rate $\beta$\cite{DEHP}.
Several steady state properties of this system are known\cite{DE}. The open boundary conditions 
in ASEP means special growth rules at the boundaries of the interface model. Thus, many
results from the ASEP can then be used to predict the properties of the interface
model with special boundary conditions\cite{DE}. In particular, for $\alpha\le 1$, $\beta\le 1$
and on the line $\alpha+\beta=1$, ASEP is known to have a factorized steady state, i.e.,
the $\tau_i$'s become independent from site to site, each with a bivariate
distribution $p(\tau)= \alpha \delta_{\tau,1} + \beta \delta_{\tau, 0}$. This
means that the slope variables in the corresponding interface model also get
uncorrelated in the stationary state. The stationary profile in space is then
described by a one dimension random walk with drift, $h_{i+1}-h_i= \xi_i$ where
$\xi_i$'s are i.i.d variables which take the value $1$ (with probability $\beta=1-\alpha$)
and $0$ (with probability $\alpha$). Thus, we have a random walk with a drift $\mu=1-2\alpha$.
For the zero drift case, $\alpha=\beta=1/2$, we would then expect that the F-Airy distribution
function, upto a scale factor,  would describe the MRH distribution of the interface in the continuum 
limit. It would be interesting to extend the method presented here to calculate the
MRH distribution for $\mu\ne 0$.

\section{ Maximal Relative Height Distribution of interfaces in the Growing regime}

In this section, we discuss the MRH distribution of a $(1+1)$-dimensional interface in its growing regime, i.e., when the 
time $1<<t<< L^z$ where $z$ is the dynamical exponent 
associated with the growth. The MRH distribution in 
the growing regime
turns out to be more universal than its stationary counterpart and holds for generic interfaces. 
Logically, one would have preferred to discuss the growing regime (early time) before the
stationary regime (late time). However, we will need to use some of the results for the stationary 
regime derived in Section IV in our discussion of the MRH distribution in the growing regime.
Hence is the reversal of the order. The MRH distribution in the growing
regime has been discussed previously in Ref.\cite{RCPS} and we will be using a similar line of arguments. Nevertheless,
we decided to include this section partly because our results would be more precise and partly to make this
paper self-contained and complete.   

Let us consider a growing interface starting from a flat initial condition at $t=0$. Initially
the heights are completely uncorrelated. As time grows, the correlation between heights also grow. 
As discussed in Section III, at any finite time $t$, the height-height correlation function 
decays with spatial distance over a characteristic correlation length $\xi(t)$ which grows
algebraically with time, $\xi(t)\sim t^{1/z}$. 
In the growing regime when the time $1<<t<< L^z$, this correlation length
is much smaller than the system size, $\xi(t) << L$. Thus, one can break up the whole system of size $L$ into
$N=L/\xi$ blocks each of size $\xi(t)$, as shown in Fig. 8. The heights (or rather the relative heights) at two points 
belonging to two different blocks are basically uncorrelated since the distance between them is bigger than the 
correlation length $\xi(t)$. However, inside a given block, the heights are strongly correlated. Let us denote
the maximal relative height in block $i$ by $h_m(i)$. These block MRH's  are shown by the black dots in Fig. 8. 
\begin{figure}[htbp]
\epsfxsize=8cm
\centerline{\epsfbox{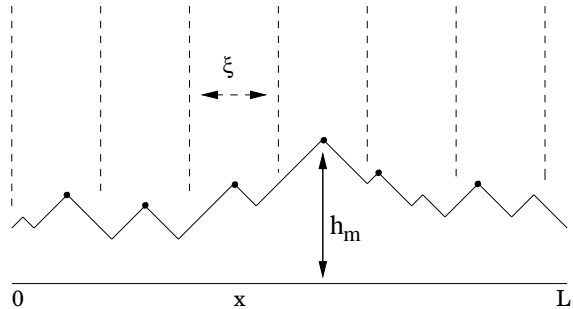}}
\caption{An interface in the growing regime with correlation length $\xi(t)$. The system is divided into $N=L/\xi(t)$ 
blocks each of length $\xi(t)$. The heights in different blocks are essentially uncorrelated. The maximal
relative height in each block is denoted by a black dot. Also shown is the global maximal relative height $h_m$.} 
\label{fig:gumbel}
\end{figure}

We are interested
in the distribution of the {\em global} MRH,
$h_m = {\rm max}\left[ h_m(1), h_m(2), \ldots, h_m(N)\right]$ where
$N$ is the number of blocks. Since the block variables $h_m(i)$'s are uncorrelated, we are thus interested in the 
extreme value statistics of a set of $N$ uncorrelated random variables, a subject that has been rather well 
studied\cite{Extreme}. Furthermore, since $\xi(t)<< L$, the system does not feel the presence of the boundaries
and there is a translational invariance over the block index $i$. This indicates that the random variables $h_m(i)$'s 
associated with different blocks are not just independent, but  
share the same distribution as well. Let us denote this common distribution by $\rho(x, \xi)= {\rm Prob}\,[h_m(i)=x, \xi]$, 
where $\xi$ is the size of the block. 
Due to the independence of different blocks, the joint distribution $P\left[\{h_m(i)\}\right]$ of the block 
variables $h_m(i)$'s factorize, $P\left[\{h_m(i)\}\right]= \prod_i \rho\left(h_m(i), \xi\right)$.

To guess the form of the distribution $\rho(x, \xi)= {\rm Prob}\,[h_m(i)=x, \xi]$, one makes a reasonable assumption 
that within a given 
block of size $\xi$, the heights have already
reached the stationary state even though the full system is still in the growing regime. Then $\rho(x)$ is just given by
the stationary MRH distribution with a system size $\xi$ and one can then directly use the results obtained in the previous 
section, namely the scaling form, $\rho(x, \xi)= {\xi}^{-1/2} W\left(x\,{\xi}^{-1/2}\right)$.
More generally, one should use the scaling form, $\rho(x, \xi)= {\xi^{-\alpha}} W\left(x\,{\xi^{-\alpha}}\right)$, where
$\alpha$ is the roughness exponent of the surface. For the $(1+1)$-dimensional EW and the KPZ equation, one has
$\alpha=1/2$. The roughness exponent $\alpha$ is related to the dynamical exponent $z$ and the growth exponent $\beta$
via the simple scaling law, $\alpha=\beta z$\cite{HZ}. 
Note that the scaling 
function $W(x)$ depends on the boundary condition. For the periodic boundary case, we had $W(x)=f(x)$ 
where $f(x)$ is the Airy distribution function in Eq. (\ref{fx1}). For the free boundary condition, at least
for the EW interface, 
$W(x)=F(x)$ with $F(x)$ being
the $F$-Airy distribution function in Eq. (\ref{frx1}). Note that it is not easy to guess the boundary conditions
at the edges of a block of size $\xi$. Therefore, it is difficult to know the precise form of the scaling function
$W(x)$. However, it turns out that in the scaling regime (see later), only the large $x$ behavior of $W(x)$ matters.
Now, both for the periodic (EW and KPZ) and the free boundary conditions (only EW), we have seen in Eqs. 
(\ref{pg1}) and (\ref{fg1}) that
the scaling function $W(x)$ has a Gaussian tail for large $x$, 
\begin{equation}
W(x)\sim \exp\left[-b x^2\right],
\label{gg1}
\end{equation}
where the constant $b=6$ for the periodic case and $b=3/2$ for the free case. This suggests that the large $x$ tail
of the scaling function is generically Gaussian, though the constant $b$ is nonuniversal and depends on the details
of the boundary conditions.

Let us first define the cumulative distribution of the global MRH in a system of size $L$ and at time $t$, 
\begin{equation}
F(h_m, L, t)={\rm Prob}\,\left[{\rm max}\{h_m(1),h_m(2),\dots, h_m(N)\}<h_m,\, L,\, t \right].
\label{gcum1}
\end{equation} 
Using the fact that the block MRH $h_m(i)$'s have a factorised joint distribution, it then follows that 
\begin{equation}
F(h_m, L, t) = {\left[\int_{-\infty}^{h_m} \rho(z, \xi)dz\right]}^N ={\left[1-\int_{h_m}^{\infty} \rho(z, \xi)dz\right]}^N.
\label{gum1}
\end{equation}
Substituting the scaling form, $\rho(x, \xi)= {{\xi}^{-\alpha}}\, W(x\,{\xi^{-\alpha}})$ in Eq. (\ref{gum1}) we get
\begin{equation}  
F(h_m, L, t) ={\left[1-\int_{ h_m/{\xi^{\alpha}} }^{\infty} W(y)dy\right]}^N\approx \exp\left[-N \int_{ h_m/{\xi^{\alpha}} 
}^{\infty} 
W(y)dy\right],
\label{gum2}
\end{equation}
where the last equality holds near the tail $h_m\to \infty$. Using the Gaussian behavior of $W(x)$ for large $x$ in Eq. 
(\ref{gg1}), it is easy to analyze the integral inside the exponential in Eq. (\ref{gum2}) and one arrives at
the limiting distribution, 
\begin{equation}
F(h_m, L, t)\to V\left(c\, \frac{D_L}{\xi^{2\alpha}}\, \left( h_m- D_L\right )\right), \quad {\rm with}\quad 
V(y)=\exp\left[-e^{-y}\right],
\label{gum3}
\end{equation}
where $D_L= \xi^{\alpha}\sqrt{\log(L/{\xi})}$, $c$ is
an unimportant nonuniversal constant and the scaling function $V(y)$ is the celebrated Gumbel function.
The probability distribution of the global MRH is then obtained by taking the derivative in Eq. (\ref{gum3}),
$P(h_m, L, t)= {\partial F(h_m, L,t)}/{\partial h_m}$. This distribution is clearly peaked around its
average value, 
\begin{equation}
\langle h_m(t) \rangle \sim \xi^{\alpha} \sqrt{\log (L/{\xi})}\sim t^{\beta}\left[{\log (L\, t^{-1/z})}\right]^{1/2}.
\label{gum4}
\end{equation}
The width of the peak around this average can also be read off Eq. (\ref{gum3})
\begin{equation}
w= \sqrt{\langle h_m^2\rangle- {\langle h_m \rangle}^2}\sim \frac{\xi^{2\alpha}}{{D_L}}\sim t^{\beta}{\left[\log 
(L\, 
t^{-1/z})\right]}^{-1/2}.
\label{gum5}
\end{equation}  
For the $(1+1)$-dimensional EW equation, $z=2$, $\beta=1/4$ and for the KPZ equation, $z=3/2$, $\beta=1/3$.
The result for the average in Eq. (\ref{gum4}) was also obtained in Ref. \cite{RCPS}, though the power $1/2$ of the 
logarithm in Eq. (\ref{gum4}) was found only numerically. In our case, this power $1/2$ is a direct consequence of
the large $x$ Gaussian tail of the scaling function $W(x)$ in Eq. (\ref{gg1}), which were derived analytically
for the EW and the KPZ equation in Section-IV and V.
Thus, the main conclusion of this section is that in 
the growing regime, the appropriately scaled global MRH has the universal Gumbel distribution that does not depend on
the details of the interface.

\section{Conclusion}

To summarize, in this paper we have studied the distribution of the global maximum $h_m$ of relative heights (,i.e., the 
height measured with respect to the spatially averaged height) in $(1+1)$-dimensional fluctuating interfaces, both of
the Edwards-Wilkinson and the Kardar-Parisi-Zhang variety. The distribution $P(h_m,L,t)$, at time $t$ and in a system 
of size $L$, has two types of scaling behaviors depending on whether one is in the {\em growing regime} (where $1<<t<<L^z$)
or in the {\em stationary regime} (where $t>> L^z$), $z$ being the dynamical exponent of the interface. In the growing 
regime, the distribution (appropriately scaled) is described by the universal Gumbel law of the extreme value statistics.
On the other hand, in the stationary regime, the distribution becomes time independent and has the scaling form,
$P(h_m, L)= L^{-1/2}W\left( h_m L^{-1/2}\right)$. The scaling function $W(x)$ depends on the boundary condition.
For the periodic boundary condition, we have analytically shown that $W(x)=f(x)$ where 
$f(x)$ is the Airy distribution
function in Eq. (\ref{fx1}) that describes the distribution of the area under a Brownian excursion over a unit interval.
In fact, in this case, the EW and the KPZ interface both share the same scaling function $f(x)$. In the case of the free 
boundary condition, we have shown that for the EW interface, $W(x)=F(x)$ where $F(x)$ is the $F$-Airy distribution function
defined in Eq. (\ref{frx1}). These analytical results have also been verified by numerical integration of the
EW and the KPZ equations. These results are summarized in Fig. 9. Since the relative heights in the stationary state
are strongly correlated, our results provide a rather rare exactly solvable case for the distribution of extremum of 
a set of {\em correlated} random variables. 
\begin{figure}[htbp]
\epsfxsize=8cm
\centerline{\epsfbox{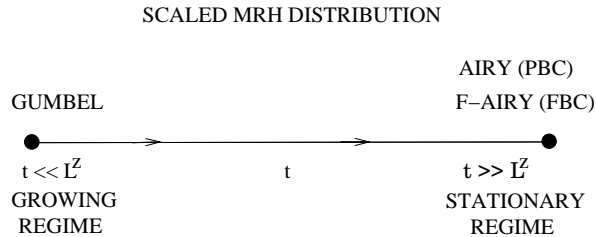}}
\caption{As time $t$ increases, the scaled MRH distribution crosses over from the universal Gumbel form
in the growing regime to either the Airy distribution function (for the periodic boundary condition (PBC)) or to the $F$-Airy 
distribution
function (for the free boundary condition (FBC)) in the stationary regime.} 
\label{fig:fp}
\end{figure}
 
Our work shows that there is yet another nontrivial scaling function in the $(1+1)$-dimensional KPZ equation, 
namely the Airy distribution function that appears in the stationary MRH distribution in $(1+1)$-dimensional 
KPZ equation with periodic boundary condition. Note that
several other nontrivial scaling functions have appeared before in $(1+1)$-dimensional KPZ equation. 
Let us list a few.

\vspace{0.1cm}

(i) In the steady state of the one dimensional EW or the KPZ equation, the
width of the surface height defined as 
\begin{equation}
w^2= \frac{1}{L}\int_0^L H^2(x,t)dx- \frac{1}{L^2}\,{\left[\int_0^L H(x,t)dx\right]}^2,
\label{stw1}
\end{equation}
was shown to have a scaling distribution, $P(w^2)\simeq {\langle w^2\rangle}^{-1}\, \Phi(w^2/\langle w^2\rangle)$
where the scaling function $\Phi(x)$ is nontrivial\cite{Width}.
 
\vspace{0.1cm}

(ii) For the $(1+1)$-dimensional KPZ equation, the probability distribution of the spatially
integrated height $Y_t=\int_0^{L} H(x,t)dx$, i.e., the area under the surface profile, was shown to 
have the following behavior at late times $t$\cite{DL}
\begin{equation}
P(Y_t) \sim \exp\left[t Z_L\left(Y_t/t\right)\right],
\label{ldv1}
\end{equation}
where $Z_L(x)$ is a large deviation function which has a Gaussian peak near $x=0$, but
has highly non-Gaussian and asymmetric tails for large $|x|$. 

\vspace{0.1cm}

(iii) More recently, it has been shown\cite{PS} that the single site height distribution 
$P(H, t)$ in the growing regime 
$t<< L^{3/2}$ of the $(1+1)$-dimensional KPZ equation scales as 
\begin{equation}
P(H,t) \sim t^{-1/3} T_W\left( \frac{H-\langle H\rangle}{t^{1/3}}\right),
\label{tw1} 
\end{equation}
where the scaling function 
$T_W(x)$ is identical (upto a scale factor) to the celebrated Tracy-Widom distribution for the largest 
eigenvalue of a random matrix drawn from the
Gaussian unitary ensemble\cite{TW}. 
Note that in the stationary regime $t>> L^{3/2}$, this single site height distribution
becomes a simple Gaussian. However, as demonstrated in this paper, the global maximum of all 
these heights in a finite system of 
size $L$ has a nontrivial distribution in the stationary regime, characterized by the Airy distribution
function for the periodic boundary condition case.
Determining the corresponding scaling function in the KPZ equation in a finite system with free 
boundary conditions remains an open problem.

In this paper we have analytically studied the distribution of the extreme height fluctuations
in the stationary state of a $(1+1)$-dimensional interface. These height fluctuations
are strongly correlated in the stationary state. Thus our work provides a rare example
where one can calculate exactly the extreme value distribution of a set of strongly
correlated random variables. In the context of interfaces, it is worth pointing out that
another type of extreme statistics has been studied recently\cite{GHPR}. In Ref. \cite{GHPR}
the authors study the distribution of the maximum of the Fourier modes of the height
fluctuations. However, in this case the Fourier modes are independent of each other
and the analytical computation is simpler, though physically relevant.

{\em Experimental Consequences}: Our work opens up an interesting experimental 
challenge to measure the Airy (or the $F$-Airy)
distribution function in a physical system. Many experimental systems are known to be well described by the 
$(1+1)$-dimensional EW equation (\ref{ew1}). Examples include, amongst others,
the high-temperature step fluctuations in crystal
surfaces\cite{Maryland,Giesen} and the displacements of
nonmagnetic particles in dipolar chains
at low magnetic field\cite{THF}. In fact, in the former system, there have been beautiful recent measurements
of a variety of interface properties such as various first-passage properties\cite{Maryland}, which were studied
before only theoretically\cite{Pers}. Hence, it would be interesting to see if the MRH distribution can also be measured
in these experimental systems. This would then be the first experimental measurement of the Airy distribution
function which has so far appeared only in mathematical problems.

Several new directions and interesting open questions emerge from this work. One of the challenges is to 
calculate the MRH distribution for interfaces in higher dimensions. For example, let us 
consider the $(d+1)$-dimensional 
EW equation, where the Laplacian term in the EW equation is $d$-dimensional. This equation for
$d>2$ describes several interesting systems, e.g., a fluctuating random Gaussian manifold and also
the time evolution of the magnetization of a spin system within mean field theory (for $d>4$).
It is simpler to continue to think in terms of interface heights growing over a $d$-dimensional
substrate.
The relative heights will again reach
a stationary joint distribution function in the long time limit,
\begin{equation}    
P\left[ \{h(\vec r)\}\right]\propto e^{-\frac{1}{2}\int {(\nabla h)}^2 d\vec r}\, \delta\left[\int h(\vec r) d{\vec r} 
\right].
\label{con1}
\end{equation}
The calculation of the distribution of the maximal relative height $h_m$, starting from the above 
joint distribution, is 
a theoretical challenge. For $d<2$, the surface will be rough in the stationary state and one expects that
the MRH will scale as, $h_m \sim L^{\alpha}$ where $\alpha=1-d/2$ is the roughness exponent and 
$L$ is the linear size
of the substrate. Hence, the MRH
distribution is expected to have the scaling form, 
$P(h_m, L)\sim L^{-\alpha} f_d \left( h_m L^{-\alpha}\right)$ where
the scaling function $f_d(x)$ will depend explicitly on the dimension $d$. On the other hand, for $d>2$, the surface
is smooth and $h_m \sim O(1)$ even in the thermodynamic limit $L\to \infty$. Hence, $P(h_m, L)$ will approach
a limiting distribution $P(h_m)$ in the thermodynamic limit. It would be interesting to compute this limiting distribution
for $d>2$. This is nontrivial because the relative heights are still correlated (the correlation function falls 
off only as a power law $1/r^{d-2}$), even though the surface is smooth.
The only simplification happens in the limit of $d\to \infty$, where the correlation function dampens exponentially
and one would expect, using the theory of extreme value statistics of independent random variables as in Sec. VI, a Gumbel 
distribution for $P(h_m)$. In fact, this has recently been verified for the EW interface defined on a small
world network\cite{GK}. However, for finite $d>1$, the problem is wide open.
The path integral techniques used in this paper are suitable only in one dimension. Entirely new techniques are required to
deal with the problem in higher dimensions. Developments of such techniques are more than welcome.

\appendix
\section{Derivation of two identities}
In this appendix, we prove two identities,
\begin{eqnarray}
\sum_{k=1}^{\infty} \frac{B(\alpha_k)}{(\alpha_k+u)}& =& \frac{\int_u^{\infty} Ai(z)dz}{Ai(u)} \label{a1} \\
\sum_{k=1}^{\infty} \frac{B^2(\alpha_k)}{(\alpha_k+u)^2} &=& {\left[\frac{\int_u^{\infty} Ai(z)dz}{Ai(u)}\right]}^2 .
\label{a2}
\end{eqnarray}
where $\alpha_k$'s are the magnitude of the zeros of the Airy function $Ai(z)$ on the negative real axis, i.e.,
$Ai(-\alpha_k)=0$ and $B(\alpha_k)$ is given by,
\begin{equation}
B(\alpha_k)= \frac{\int_{-\alpha_k}^{\infty} Ai(z)dz}{Ai'(-\alpha_k)},
\label{a3}
\end{equation}
where $Ai'(z)=dAi(z)/dz$. The first identity was used in Eq. (\ref{marea6}). The second identity appeared in Eq. (\ref{flt3})
where $C(\alpha_k)=B^2(\alpha_k)$.

To prove the first identity in Eq. (\ref{a1}), we consider the following contour integral in the complex $z$ plane,
\begin{equation}
I_1= \int_C \frac{dz}{2\pi i}\, \frac{1}{(z-u)}\, \frac{\int_z^{\infty} Ai(z')dz'}{Ai(z)},
\label{a4}
\end{equation}
where the contour $C$ is shown in Fig. 10. Note that the integrand in Eq. (\ref{a4}) has simple poles
at $z=u$ and $z=-\alpha_k$ for $k=1,2,\ldots$. The latter fact follows since $-\alpha_k$ is a simple zero of
$Ai(z)$.
\begin{figure}[htbp]
\epsfxsize=8cm
\centerline{\epsfbox{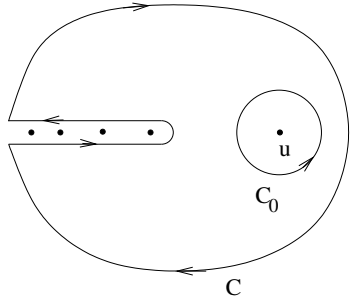}}
\caption{The contour for the integral in Eq. (\ref{a4}). The dots on the negative real axis denote the zeros
$-\alpha_k$'s of the Airy function $Ai(z)$.}
\label{fig:contour}
\end{figure}

Using the asymptotic expansion of $Ai(z)\sim z^{-1/4}\exp[-2z^{2/3}/3]$ for large $z$\cite{AS}, it is easy to
see that the function $Q(z)= \int_z^{\infty} Ai(z')dz'/Ai(z) \sim z^{-1/2}$ for large $z$. Hence, once we extend
the radius of the contour $C$ to infinity, the contribution to the integral $I_1$ in Eq. (\ref{a4}) from the circular part
of the contour tends to zero. The only remaining contribution comes from the arms of the contour around the
negative real axis, which is equal to the sum of the residues around $-\alpha_k$'s each of which is a simple pole
of the function $Q(z)$. Evaluating the residues around these poles in Eq. (\ref{a4}) we get
\begin{equation}
I_1= - \sum_{k=1}^{\infty} \frac{B(\alpha_k)}{(\alpha_k+u)}.
\label{a5}
\end{equation}
On the other hand, since the integrand in Eq. (\ref{a4}) is analytic 
in the interior of the domain between the contour $C$ and 
$C_0$ in Fig. 10, it follows that $\int_C = -\int_{C_0}$. But the integral around $C_0$ is just the residue at the 
pole $z=u$. Evaluating this residue, we then get
\begin{equation}
I_1= - \frac{\int_u^{\infty} Ai(z)dz}{Ai(u)}
\label{a6}
\end{equation}
Comparing Eqs. (\ref{a5}) and (\ref{a6}), we get the identity in Eq. (\ref{a1}). 

The identity in Eq. (\ref{a2}) can be proved exactly in the same manner. We now consider the contour 
integral,
\begin{equation}
I_2= \int_C \frac{dz}{2\pi i}\, \frac{1}{(z-u)}\, {\left[\frac{\int_z^{\infty} Ai(z')dz'}{Ai(z)}\right]}^2,
\label{a7}
\end{equation}
around the same contour $C$ in Fig. 10. Note that the integrand in Eq. (\ref{a7}) now has doubles poles at $z=
-\alpha_k$ for all $k=1,2,\ldots$ (apart from the single pole at $z=u$). The rest of the calculation
is similar as in the evaluation of $I_1$ and we do not repeat them here. Following these steps
one easily arrives at the identity in Eq. (\ref{a2}).

\section{The Inversion of a Laplace Transform}

In this appendix, we wish to express the function $F(x)$ explicitly in real space $x$ by inverting its Laplace transform,
\begin{equation}
\int_0^{\infty} F(x)\, e^{-sx} dx= 2^{-1/3} s^{2/3}\, \sum_{k=1}^{\infty} C(\alpha_k)\,  e^{-\alpha_k s^{2/3} 2^{-1/3}},
\label{b1}
\end{equation}
where the constant, $C(\alpha_k)={\left[B(\alpha_k)\right]}^2$ with $B(\alpha_k)$ given in Eq. (\ref{a3}). 

To proceed, we use an identity originally derived by Takacs in the context of Brownian excursions\cite{Takacs},
\begin{equation}
2^{1/3} \int_0^{\infty} J\left(2^{1/3}\, x^{-2/3}\right) x^{-5/3} e^{-s x} dx = e^{-s^{2/3}},
\label{b2}
\end{equation}
where the function $J(x)$ can be expressed in terms of the known hypergeometric function $U(a,b,z)$,
\begin{equation}
J(x)= \frac{2^{2/3}x}{3^{3/2} \sqrt{\pi}}\, U\left(1/6, 4/3, 2x^3/{27}\right)\, e^{-2x^3/{27}}.
\label{b3}
\end{equation}
Thus the identity in Eq. (\ref{b2}) gives us the inverse Laplace transform,
\begin{equation}
{\cal L}_s^{-1}\left[e^{-s^{2/3}}\right]= 2^{1/3} x^{-5/3} J\left(2^{1/3}\, x^{-2/3}\right)
\label{b4}
\end{equation}
where $J(x)$ is given by Eq. (\ref{b3}).

Our first step is to rescale $s$ in Eq. (\ref{b4}) by a constant factor, i.e.,
$s\to \gamma^{3/2} s$. This is equivalent to the rescaling, $x\to \gamma^{-3/2} x$. Then, Eq. 
(\ref{b4}) transforms into
\begin{equation}
{\cal L}_s^{-1}\left[e^{-\gamma s^{2/3}}\right]= \gamma\, 2^{1/3} x^{-5/3} J\left(\gamma 2^{1/3} x^{-2/3}\right).
\label{b5}
\end{equation}
Differentiating Eq. (\ref{b5}) with respect to $\gamma$ gives
\begin{equation}
{\cal L}_s^{-1}\left[2^{-1/3} s^{2/3} e^{-\gamma s^{2/3}}\right]= - x^{-5/3} J\left(2^{1/3}\, x^{-2/3}\right)-\gamma 2^{1/3} 
x^{-7/3} J'\left(\gamma 2^{1/3} x^{-2/3}\right),
\label{b6}
\end{equation}
where $J'(x)=dJ/dx$. We take the derivative of the function $J(x)$ in Eq. (\ref{b3}) and use the following identity\cite{AS}
satisfied by the function $U(a,b,x)$
\begin{equation}
x \frac{\partial U(a,b,x)}{\partial x}= (a-b+x)\, U(a,b,x)- U(a-1,b, x).
\label{b7}
\end{equation}
Rearranging and simplifying, we get
\begin{equation}
{\cal L}_s^{-1}\left[2^{-1/3} s^{2/3} e^{-\gamma s^{2/3}}\right]= \frac{\gamma}{\sqrt{3\pi}\, x^{7/3}}\,
\left[ U\left(1/2, 4/3, 4\gamma^3/{27 x^2}\right) + 2 U\left(-5/6, 4/3, 4\gamma^3/{27 x^2}\right)\right]\,e^{-4\gamma^3/{27 
x^2}}.
\label{b8}
\end{equation}
Note that this formula in Eq. (\ref{b8}) is exactly what we need to invert the Laplace transform in Eq. (\ref{b1}). 
Substituting $\gamma= 2^{-1/3}\alpha_k$ in Eq. (\ref{b8}) and using this formula in Eq. (\ref{b1}) we get
\begin{equation}
F(x)= \frac{2^{-1/3}}{{\sqrt{3\pi}}\, x^{7/3}}\sum_{k=1}^{\infty} \alpha_k C(\alpha_k)\left[U\left(1/6,4/3,b_k/x^2\right)+ 2
U\left(-5/6, 4/3,
b_k/x^2\right)\right] e^{-b_k/x^2}
\label{b9}
\end{equation}
where $b_k= 2\alpha_k^3/{27}$.



\begin{thebibliography}{999}

\bibitem{Darling} D.A. Darling, On the supremum of certain Gaussian processes, 
{\em Ann. Probab.} {\bf 11}:803-806 (1983).

\bibitem{Louchard} G. Louchard, Kac's formula, Levy's local time and Brownian
excursion, {\em  J. Appl. Prob.} {\bf 21}:479-499 (1984). 

\bibitem{Takacs} L. Takacs, A Bernoulli excursion and its various
applications, {\em  Adv. Appl. Prob.} {\bf 23}:557-585, (1991); Limit distributions
for the Bernoulli meander, {\em  J. Appl. Prob.} 
{\bf 32}:375-395 (1995).

\bibitem{AS} M. Abramowitz and I.A. Stegun, {\em Handbook of Mathematical Functions}
(Dover, New York, 1973).

\bibitem{CSY} M. Cs\"org\"o, Z. Shi, and M. Yor, Some asymptotic properties
of the local time of the uniform empirical processes, {\em  Bernoulli.} 
{\bf 5}:1035-1058 (1999).

\bibitem{FPV} P. Flajolet, P. Poblete, and A. Viola, On the analysis of
linear probing hashing, {\em  Algorithmica.} 
{\bf 22}:490-515 (1998).

\bibitem{FL} P. Flajolet and G. Louchard, Analytic variations on the Airy
distribution, {\em Algorithmica.} {\bf 31}:361-377 (2001).

\bibitem{FSS} P. Flajolet, B. Salvy, and G. Schaeffer, Airy phenomena and
analytic combinatorics of connected graphs, {\em Electronic J. of
Combinatorics} {\bf 11}:1-30 (2004).

\bibitem{MR} C.L. Mallows and J. Riordan, The inversion enumerator for
labelled trees, {\em  Bull. Am. Math. Soc.} 
{\bf 74}:92-94 (1968); I. Gessel, B.E. Sagan, and 
Y.-N. Yeh, Enumeration of trees by inversion, {\em  J. Graph Theory} {\bf 19}:435-459 (1995).

\bibitem{Wright} E.M. Wright, The number of connected sparsely edged
graphs, {\em
J. Graph Theory} {\bf 1}:317-330 (1977); 2- Smooth graphs and blocks {\bf 2}:299-305 (1978); 3- Asymptotic results {\bf 4}, 393-407 (1980).

\bibitem{FKP} P. Flajolet, D.E. Knuth, and B. Pittel, The first cycles in an evolving graph, {\em Discrete 
Math.}
{\bf 75}:167-215 (1989); S. Janson, D.E. 
Knuth, T. Luczak, and B. Pittel, {\em Random Structures and Algorithms} 
{\bf 4}:233 (1993).

\bibitem{RGJ} C. Richard, A.J. Guttmann, and I. Jensen, Scaling function and
universal amplitude combinations for self avoiding polygons, {\em J. Phys. A: Math. Gen.} {\bf 34}:L495-501 (2001); C. 
Richard, Scaling behaviour of the two-dimensional polygons models, {\em J. Stat. Phys.} {\bf 108}:459-493 (2002); C. Richard, I. Jensen, 
and A.J. Guttmann, Scaling function for self-avoiding polygons, {\em Proceedings of 
the International Congress on Theoretical Physics, Paris, July 2002, ed. by D. Iagolnitzer, D. Rivasseau, and
J. Zinn-Justin} (Birkhauser), cond-mat/0302513.

\bibitem{R1} C. Richard, Area distribution of the planar random loop
boundary, {\em J. Phys. A: Math. Gen.} {\bf 37}:4493-4500 (2004).

\bibitem{MC1} S.N. Majumdar and A. Comtet, Exact maximal height distribution
of fluctuating interfaces, {\em Phys. Rev. Lett.} {\bf 92}:225501 (2004).

\bibitem{Review} A.L. Barabasi and H.E. Stanley, {\em Fractal Concepts in Surface
Growth} (Cambridge University Press, Cambridge, England, 1995); J. Krug, {\em Adv. Phys.}
{\bf 46}:139-282 (1997).

\bibitem{HZ} T. Halpin-Healy and Y.-C. Zhang, Kinetic roughening phenomena,
stochastic growth, directed polymers and all that, {\em  Phys. Rep.} {\bf 254}:215-414 (1995).

\bibitem{Extreme} E.J. Gumbel, {\em Statistics of Extremes} (Columbia University Press, New York,
1958).

\bibitem{DS} J.-P. Bouchaud and M. M\'ezard, Universality classes for extreme
value statistics, {\em J. Phys. A} {\bf 30}:7997-8015 (1997); D. Carpentier and P. Le
Doussal, Glass transition of a particle in a random potential, front selection
in non-linear renormalization group and entropic phenomena in Liouville and
sinh-Gordon models, {\em  Phys. Rev. E} {\bf 63}:026110 (2001); D.S. Dean and S.N.
Majumdar, Extreme-value statistics of hierarchically correlated variables,
deviation from Gumbel statistics and anomalous persistence, {\em Phys. Rev. E} {\bf 64}:046121 (2001); 
P. LeDoussal and C. Monthus, Exact solutions for the
statistics of extrema of some random 1D landscapes, application to the
equilibrium and the dynamics of the toy model, {\em Physica A} {\bf 317}:140-198 (2003).

\bibitem{Trees} For a review see, S.N. Majumdar and P.L. Krapivsky, Extreme value statistics and traveling
fronts: various applications, {\em Physica A} {\bf 318}: 161-170 (2003).

\bibitem{Net} For a review see, E. Ben-Naim, P.L. Krapivsky, and S.
Redner, extremal properties of random structures,
cond-mat/0311552.

\bibitem{RCPS} S. Raychaudhuri, M. Cranston, C. Przybyla, and Y.
Shapir, Maximal height scaling of kinetically growing surfaces, {\em Phys. Rev.
Lett.} {\bf 87}:136101 (2001).

\bibitem{Edwards} S.F. Edwards and D.R. Wilkinson, The surface statistics of
a granular agregate, {\em  Proc. R. Soc. London A} {\bf 381}:17-31 (1982).

\bibitem{KPZ} M. Kardar, G. Parisi and Y.-C. Zhang, Dynamical scaling of
growing interfaces, {\em Phys. Rev. Lett.} {\bf 56}:889-892 (1986).

\bibitem{PW} M. Perman and J.A. Wellner, On the distribution of Brownian
areas, {\em  Ann. Appl. Prob.} {\bf 6}:1091-1111 (1996).

\bibitem{Jeanblanc} M. Jeanblanc, J. Pitman and M. Yor, The Feynman-Kac
formula and decomposition of Brownian paths, {\em Comput. and Appl. Math.}
{\bf 16}:27-52 (1997).

\bibitem{NThe} M. Nguyen Th\^e, Area and inertial moments of Dyck paths, Combinatorics, Probability and 
Computing (2003), submitted.

\bibitem{Redner} S. Redner, {\em A Guide to First-passage Processes}
(Cambridge University Press, Cambridge 2001).

\bibitem{Albright} J.R. Albright, Integrals of products of 
Airy functions, {\em J. Phys. A} {\bf 10}:485-490 (1977).

\bibitem{Burkhardt} T.W. Burkhardt, Semiflexible polymer in the half plane
and statistics of the integral of a Brownian curve, {\em J. Phys. A} {\bf 26}:L1157-1162 (1993).

\bibitem{Width} G. Foltin, K. Oerding, Z. Racz, R.L. Workman, and R.K.P. Zia,
Width distribution for random-walk interfaces, {\em  Phys.
Rev. E} {\bf 50}:R639-642 (1994); Z. Racz and M. Plischke, Width distribution
for 2+1 dimensional growth and deposition processes, {\em  Phys. Rev. E} {\bf
50}:3530-3537 (1994).
 
\bibitem{Rouse} A. Yu. Grosberg and A.R. Khokhlov, {\em Statistical physics of
macromolecules} (AIP Press, 1994).

\bibitem{Feller} W. Feller, {\em Introduction to Probability Theory and its 
Applications}, 3rd ed. (Wiley, New York, 1968), Vol. I.

\bibitem{AF} J. Amar and F. Family, Numerical solution of continuum equation
for interface growth in 2+1 dimension, {\em Phys. Rev. A} {\bf 41}:3399-3402 (1990).

\bibitem{MKD} K. Moser, J. Kert\'esz, and D.E. Wolf, Numerical solution of
the Kardar-Parisi-Zhang equation in one,two and three dimensions, {\em Physica A} {\bf 178}:215-226 (1991); 
K. Moser and D.E. Wolf, Vectorized and parallel simulations of the
Kardar-Parisi-Zhang equation in 3+1 dimensions, {\em  J. Phys. A} {\bf 27}:4049-4054 (1994).

\bibitem{DDK} C. Dasgupta, S. Das Sarma, and J.M. Kim, Controlled instability
and multiscaling in models of epitaxial growth, {\em  Phys. Rev. E} {\bf 54}:R4552-4555 (1996); 
C. Dasgupta, J.M. Kim, M. Dutta, and S. Das Sarma,
Instability, intermittency, and multiscaling in discrete growth models of
kinetic roughening, ibid. {\bf 55}:2235-2254 (1997).

\bibitem{NB} T.J. Newman and A.J. Bray, Strong coupling behaviour in discrete
Kardar-Parisi-Zhang equation, {\em  J. Phys. A} {\bf 29}:7917-7928 (1996).

\bibitem{LS1} C.-H. Lam and F.G. Shin, Anomaly in numerical integration of
the Kardar-Parisi-Zhang equation, {\em Phys. Rev. E} {\bf 57}:6506-6511 (1998).

\bibitem{LS2} C.-H. Lam and F.G. Shin, Improved discretisation of the
Kardar-Parisi-Zhang equation, {\em Phys. Rev. E} {\bf 58}:5592-5595 (1998).

\bibitem{KS2} See Eq. (5.26) in Ref. \cite{KS} where this
result was attributed to unpublished work (1989) of T. Nieuwenhuizen.

\bibitem{KS} For a review, see J. Krug and H. Spohn, in {\em Solids Far From
Equilibrium: Growth, Morphology and Defects}, ed. by C. Godreche (Cambridge
University Press, Cambridge, 1991).

\bibitem{Sstep} P. Meakin et. al.,  Ballistic deposition on surfaces,  
{\em Phys. Rev. A} {\bf 34}, 5091 (1986). 

\bibitem{DEHP} B. Derrida et. al., Exact solution of a $1$-d asymmetric exclusion model using a 
matrix formulation,  J. Phys. A {\bf 26}, 1493 (1993).

\bibitem{DE} For a review, see B. Derrida and M.R. Evans in {\em Nonequilibrium
Statistical Mechanics in One Dimension}, ed. by V. Privman (Cambridge University
Press, Cambridge, 1997).

\bibitem{DL} B. Derrida and J.L. Lebowitz, Exact large deviation function in
the asymmetric exclusion process, {\em Phys. Rev. Lett.} {\bf 80}:209-213 (1998);
B. Derrida and C. Appert, Universal large deviation function of the
Kardar-Parisi-Zhang equation in one dimension, {\em J. Stat. Phys.} {\bf 94}:1-30 (1999).

\bibitem{PS} M. Praehofer and H. Spohn, Universal distribution of growth
processes in 1+1 dimensions and random matrices, {\em Phys. Rev. Lett.} {\bf
84}:4882-4885 (2000);
K. Johansson, Shape fluctuations and random matrices {\em  Comm. Math. Phys.} {\bf 209}:437-476 (2000);
J. Gravner, C.A. Tracy and H. Widom, Limit theorems for height fluctuations in a class of discrete space 
and time growth models {\em J. Stat. Phys.} {\bf 102}:1085-1132 (2001);
S.N. Majumdar and S. Nechaev, Anisotropic ballistic deposition model with
links to the Ulam problem and the Tracy-Widom distribution, {\em Phys. Rev. E }
{\bf 69}:011103 (2003).

\bibitem{TW} C.A. Tracy and H. Widom, Level-spacing distributions and the
Airy kernel, {\em Commun. Math. Phys.} {\bf 159}:151-174 (1994).

\bibitem{GHPR} G. Gyorgyi et. al., Statistics of extreme intensities for Gaussian 
interfaces, {\em Phys. Rev. E} {\bf 68}, 056116 (2003).

\bibitem{Maryland} D.B. Dougherty et. al, Experimental persistence
probability for fluctuating steps, {\em  Phys. Rev. Lett.} {\bf 89}:136102-136106 (2002);
M. Constantin et. al, Infinite family of persistence exponents for interface
fluctuations, {\em  Phys. Rev. Lett.} {\bf 91}:086103 (2003);
C. Dasgupta et. al, Survival in equilibrium step fluctuations, {\em Phys. Rev. E} {\bf 69}: 022101 (2004).

\bibitem{Giesen} M. Giesen, Step and island dynamics at solid/vacuum and solid/liquid interfaces {\em Prog. 
Surf. Sci.} {\bf 68}:1-153 (2001) and references therein.

\bibitem{THF} R. Toussaint, G. Helgesen, and E.G. Flekkoy, Dynamic roughening
and fluctuations of dipolar chains, cond-mat/0311340.

\bibitem{Pers} J. Krug et. al, Persistence exponents for fluctuating
interfaces, {\em  Phys. Rev. E} {\bf 56}:2702-2712 (1997); H. Kallabis and J.
Krug, Persistence of Kardar-Parisi-Zhang interfaces {\em Europhys. Lett.} {\bf 45}:20-25 (1999);
Z. Toroczkai and E.D. Williams, Nanoscale fluctuations at solid interfaces, {\em
Physics Today} {\bf 52}:No. 12, 24-28 (1999);
S.N. Majumdar and A.J. Bray, Spatial persistence of fluctuating interfaces, {\em Phys. Rev. Lett.} 
{\bf 86}:3700-3703 (2001). J. Krug, Power law in surface physics: the deep, the shallow
and the useful, Physica A {\bf 340}, 647 (2004).

\bibitem{GK} H. Guclu and G. Korniss, Extreme fluctuations in small-world
networks with relaxional dynamics, {\em  Phys. Rev. E} {\bf 69}:065104(R) (2004).


\end{thebibliography}
\end{document}